\documentclass[11pt]{article}
\pdfoutput=1
\usepackage{cite}
\usepackage{booktabs}
\usepackage[english]{babel}
\usepackage{mathtools}
\usepackage{amsmath,amssymb,amsbsy,amstext, amsthm, simplewick, amsfonts}
\usepackage{hyperref}
\usepackage{graphicx}
\usepackage{amsfonts}
\usepackage{amssymb}
\usepackage[small]{caption}
\usepackage{upgreek}
\usepackage[svgnames,dvipsnames,x11names,table]{xcolor}
\usepackage{multirow}
\usepackage{geometry}
\usepackage[hang,flushmargin]{footmisc}
\usepackage{bm}
\usepackage{braket}
\usepackage{subcaption}
\usepackage{mathtools}
\usepackage{setspace}
\usepackage{cleveref}
\usepackage{comment}
\usepackage{scalerel}
\usepackage[normalem]{ulem}
\usepackage{slashed}
\usepackage{tensor}
\usepackage{enumitem}
\usepackage{ifthen}
\usepackage{appendix}
\usepackage{floatrow}
\usepackage{siunitx}
\usepackage{framed}
\usepackage{float}
\usepackage{fontawesome}
\usepackage{geometry}
\usepackage{soul}
\usepackage{cancel}
\usepackage{tikz}
\usetikzlibrary{decorations.markings}
\usetikzlibrary{shapes.misc}
\usetikzlibrary{arrows}
\usetikzlibrary{fit, chains}

\makeatletter
\g@addto@macro\bfseries{\boldmath}
\makeatother

\numberwithin{equation}{section} 

\newcolumntype{M}[1]{>{\centering\arraybackslash}m{#1}}
\tikzset{arrow/.style = {thick,->,>=stealth,}}
\tikzset{nearnodes/.style={node distance = 1cm,}}
\tikzset{model1/.style = {rectangle, text width = 3.8cm, minimum height=6.2cm, text centered, draw = black, fill=gray!10,}}

\definecolor{pyblue}{RGB}{31, 119, 180}
\definecolor{pyorange}{RGB}{255, 127, 14}
\definecolor{pygreen}{RGB}{44, 160, 44}
\definecolor{pyred}{RGB}{214, 39, 40}
\definecolor{lightgray}{gray}{0.9}

\hypersetup{
    colorlinks=true,
    linkcolor={Purple!80!black},
    citecolor={pyblue},
    urlcolor={Purple!80!black}
}

\usepackage{colortbl}

\makeatletter
\newlength{\apb@width}
\newcommand{\autoparbox}[2][c]{\settowidth{\apb@width}{#2}\parbox[#1]{\apb@width}{#2}}

\makeatother

\usepackage{listings}
\usepackage{color}

\usepackage[framemethod=default]{mdframed}
\newmdenv[skipabove=7pt,
skipbelow=7pt,
rightline=false,
leftline=false,
topline=false,
bottomline=false,
backgroundcolor=gray!10,
linecolor=gray,
innerleftmargin=5pt,
innerrightmargin=5pt,
innertopmargin=5pt,
innerbottommargin=5pt,
leftmargin=0cm,
rightmargin=0cm,
linewidth=4pt]{eBox}

\newcommand\blfootnote[1]{%
  \begingroup
  \renewcommand\thefootnote{}\footnote{#1}%
  \addtocounter{footnote}{-1}%
  \endgroup
}

\usepackage[most]{tcolorbox}
\tcbset{colback=white, colframe=black,
        highlight math style= {enhanced,
            colframe=red,colback=red!10!white,boxsep=0pt}
        }

\crefname{table}{Table}{Tables}
\crefname{equation}{Eq.}{Eqs.}
\crefname{appendix}{App.}{Apps.}
\crefname{section}{Section}{Secs.}
\crefname{figure}{Fig.}{Figs.}

\def \d {\mathrm{d}}

\def \I {\mathcal{I}}
\def \O {\mathcal{O}}
\def \k {\bm{k}}
\def \Re {\mathcal{R}}
\def \Im {\mathcal{I}}

\begin{document}

\begin{titlepage}
\setcounter{page}{1} \baselineskip=15.5pt
\thispagestyle{empty}
$\quad$
\vskip 0 pt

\vspace*{0cm}

\begin{center}
{\fontsize{25}{18} \bf \textcolor{pyblue}{$\mathcal{C}$}osmo\textcolor{pyred}{$\mathcal{F}$}low}\\[10pt] 
{\fontsize{15}{18}  \it Python Package for Cosmological Correlators}
\end{center}

\begin{figure}[h!]
   \centering
   \includegraphics[width=0.8\textwidth]{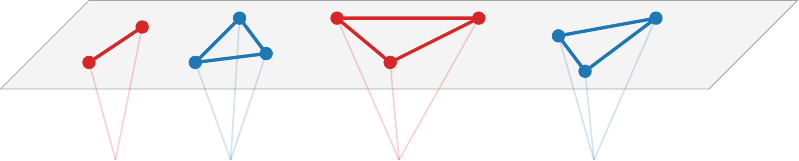}
\end{figure}

\vskip 15pt
\begin{center}
\noindent
{\fontsize{12}{18}\selectfont Denis Werth$^\dagger$\footnote{Corresponding author.}\blfootnote{\href{mailto:denis.werth@iap.fr}{denis.werth@iap.fr}}, Lucas Pinol,$^{\dagger \ddagger}$\blfootnote{\href{mailto:lucas.pinol@phys.ens.fr}{lucas.pinol@phys.ens.fr}} and Sébastien Renaux-Petel$^{\dagger}$\blfootnote{\href{mailto:renaux@iap.fr}{renaux@iap.fr}}}
\end{center}

\begin{center}
\textit{$^\dagger$ Institut d'Astrophysique de Paris, Sorbonne Université, CNRS, Paris, F-75014, France}

\vskip 5pt
\textit{$^\ddagger$ Laboratoire de Physique de l’École Normale Supérieure, Sorbonne Université,\\ Université Paris Cité, CNRS, Paris, F-75005, France} 
\end{center}

\vspace{1cm}
\begin{center}{\bf Abstract}
\end{center}

\noindent Cosmological correlators hold the key to high-energy physics as they probe the earliest moments of our Universe, and conceal hidden mathematical structures. However, even at tree-level, perturbative calculations are limited by technical difficulties absent in flatspace Feynman diagrammatics. In this paper, we introduce \textsf{CosmoFlow}: a new accurate open source Python code that computes tree-level cosmological correlators by tracing their time flow. This code is specifically designed to offer a simple, intuitive and flexible coding environment to theorists, primordial and late-time cosmologists. It can typically serve to complement analytical computations, to provide physical intuition when studying various inflationary theories, and to obtain exact results in regimes that are analytically out of reach. This paper presents the basic structure of \textsf{CosmoFlow}, leads the reader through an in-depth user-guide, and illustrates how it can be used with a series of worked examples. Our hope is that this first building block sets the stage for a bank of theoretical data, which can be nurtured and enhanced collaboratively by the community. \textsf{CosmoFlow} is publicly available on \href{https://github.com/deniswerth/CosmoFlow}{GitHub}.

\end{titlepage}

\setcounter{page}{2}

\restoregeometry

\begin{spacing}{1.2}
\newpage
\setcounter{tocdepth}{3}
\tableofcontents
\end{spacing}

\setstretch{1.1}
\newpage

\section{Introduction}

Interpreting data from either particle collisions or cosmological observations crucially relies on a close interplay between accurate theoretical predictions and precise experiments.

\vskip 4pt
In particle physics, precision calculations are mandatory for the measurement of properties of particles, such as masses or spins, and fundamental parameters such as coupling constants. In this case, the crucial objects that map theory and observables are \textit{cross sections}, or more fundamentally the underlying scattering amplitudes. These objects contain the dynamical information associated with the models used to describe data, and can be obtained from Feynman diagrams. Perturbative calculations are usually truncated at a certain order that dictates the level of accuracy to confront against the data. Although in principle systematic, this approach hides many challenges. In fact, these calculations quickly lead to an explosion of both analytic and algebraic complexity, that grow exponentially with the order at which the perturbative expansion is truncated. 

\vskip 4pt
From the theoretical side, probing fine features requires increasingly accurate predictions, ensuring that all various possible processes are covered. For example, with the LHC experiments achieving percent-level precision measurements, QCD perturbative calculations at NNNLO become necessary, large logarithms need to be resummed in order to get reliable predictions, and multi-loop Feynman integrals for complicated kinematical processes need to be evaluated. In short, such a precision imposes enormous demands on theory. Consequently, a large number of automated tools have been developed and are widely used by the community to make these accurate predictions. Examples of numerical packages to compute higher-order diagrams include \textsf{pySecDec}~\cite{Borowka:2017idc}, \textsf{FIESTA}~\cite{Smirnov:2008py, Smirnov:2009pb, Smirnov:2013eza}, \textsf{DiffExp}~\cite{Hidding:2020ytt}, \textsf{AMFlow}~\cite{Liu:2022chg}, and \textsf{FormCalc}~\cite{Hahn:1998yk}, some of which can accommodate parallelisable computation and the usage of GPUs. A large number of events for hadronic collider physics can be generated, e.g.~with \textsf{MadGraph}~\cite{Stelzer:1994ta, Alwall:2007st, Alwall:2011uj}. Additionally, tools like \textsf{CompHEP}~\cite{Pukhov:1999gg, CompHEP:2004qpa} and \textsf{CalcHEP}~\cite{Pukhov:2004ca, Belyaev:2012qa} (see also~\cite{Alwall:2014hca}) directly accept Lagrangians as inputs and automatically compute the corresponding cross sections, with \textsf{FeynRules}~\cite{Christensen:2008py, Christensen:2009jx, Alloul:2013bka} and \textsf{LanHEP}\cite{Semenov:2008jy} deriving the underlying Feynman rules.

\vskip 4pt
In cosmology, confronting theory against observations of, e.g., the large-scale structure of the Universe, requires an accurate modelling of inhomogeneities and their time evolution. Similarly to particle physics, as long as these perturbations remain relatively small, they can be treated in perturbation theory. Systematic approaches to make predictions, for example by evolving fluctuations from early to later times, have enabled the developments of automated tools. To name a few, these tools include publicly-available Boltzmann codes such as \textsf{CAMB}~\cite{Lewis:1999bs} and \textsf{CLASS}~\cite{Blas:2011rf}, numerical implementations of the EFTofLSS such as \textsf{PyBird}~\cite{DAmico:2020kxu} to compute correlators of biased tracers in redshift space, or N-body codes like \textsf{Quijote}~\cite{Villaescusa-Navarro:2019bje} to simulate the late-time Universe where fluctuations cannot be treated perturbatively.\footnote{A non-exhaustive list of open source numerical packages/libraries/tools used in cosmology (as well as in high-energy physics and astrophysics) can be found \href{https://github.com/nikosarcevic/HEP-ASTRO-COSMO/\#cosmo}{here}.}

\vskip 4pt
Quite remarkably and under relatively mild assumptions, all observed fluctuations in the Universe can be traced back to the beginning of the Hot Big Bang, where they become spatial correlations on the end of inflation surface. These spatial correlations inherit in their structure and kinematics all the information about the unknown and potentially new physics in the bulk of the spacetime, i.e.~during inflation. As such, these \textit{cosmological correlators} are the precise analogue of the previously mentioned cross sections, as they are the crucial links connecting theory of the early Universe and observations of later-time correlated fluctuations. On large scales, primordial fluctuations are weakly coupled and similar technics as in particle physics, i.e.~perturbative QFT, are widely employed to compute correlators. Perturbative Schwinger-Keldysh diagrammatics~\cite{Weinberg:2005vy, Chen:2017ryl} offer a complete algorithmic way of writing down the expression of any cosmological correlator. However, actual computations hide a daunting complexity, absent in flat-space calculations, and the few known results are in some cases not illuminating. Even the simplest tree-level processes sometimes do not have exploitable closed form expressions. The reasons are multiple. Contrary to particle physics, interactions are not localised in time. As a result, the effect of interactions needs to be tracked and integrated over from the infinite past to the end of inflation. The evolving background leads to additional (nested) time integrals to perform, and distorts the free propagation of fields whose mode functions are no longer simple plane waves but complicated special functions or simply unknown. Numerical tools to compute cosmological correlators up to three-point functions have been developed, including e.g.~\cite{Hazra:2012yn, Sreenath:2014nca}, and the numerical implementation of the transport approach \textsf{CppTransport}/\textsf{PyTransport}~\cite{Dias:2016rjq, Seery:2016lko, Mulryne:2016mzv, Ronayne:2017qzn}. However, these methods solve at the same time the background dynamics governing the homogeneous spacetime evolution, and corresponding fluctuations whose couplings are precisely dictated by the background evolution. Consequently, these approaches operate on a case-specific basis, rendering them inadequate for handling general theories of fluctuations due to the necessity of more complicated background models. In turn, this can significantly skew our data interpretation, revealing the need to formulate an approach capable of providing accurate predictions across all physically motivated inflationary theories. Ergo, there is a need for a flexible and robust tool for computing cosmological correlators in all theories.

\vskip 4pt
In this paper, we present \textcolor{pyblue}{$\bm{\mathcal{C}}$}\textsf{osmo}\textcolor{pyred}{$\bm{\mathcal{F}}$}\textsf{low}, an open source Python package designed to compute cosmological correlators. The approach is based on following their time evolution from their origin as quantum zero-point fluctuations to the end of inflation. The present code automatically solves universal differential equations in time satisfied by equal-time correlators that dictate their bulk time evolution. In this sense, our numerical implementation of the cosmological flow~\cite{Werth:2023pfl, Pinol:2023oux} is about integrating by differentiating, and is completely equivalent to the canonical in-in Schwinger-Keldysh formalism. Our code can compute any tree-level two- and three-point correlators in any theory in all kinematic configurations. It has clear separation between the physics and the technical implementation details, and is designed so that computations can be performed with different parameters, without requiring to re-implement different theories each time. This allows the user to easily navigate through various theories. Our main credo while developing \textsf{CosmoFlow} is \textit{simplicity}. Therefore, even for basic-level users, we believe that using and editing the code, as well as implementing new theories, are elementary tasks. The code and numerous ready-to-be-executed notebooks are available on \href{https://github.com/deniswerth/CosmoFlow}{GitHub}. If you would like to extend and/or improve some aspects of \textsf{CosmoFlow}, please contact us and let us know about your ideas. The code is free of use but for any research done with it or modifications of it, you are kindly asked to cite this code paper together with the two papers that present the Cosmological Flow~\cite{Werth:2023pfl, Pinol:2023oux}.

\newpage

\paragraph{Why is \textsf{CosmoFlow} useful?} The huge complexity in computing cosmological correlators has compelled theorists to focus on rather simple theories, and correlators are typically computed under several stringent assumptions.\footnote{Even numerically, the convergence of the integrals in the UV with the $i\epsilon$ prescription has always been a long-standing problem, see e.g.~\cite{Chen:2006xjb, Chen:2008wn, Junaid:2015hga, Tran:2022euk}.} This includes for example studying single-field theories with perfect or almost-perfect scale invariance. When several fields are considered, particles are assumed to be weakly mixed, and typically only single-exchange diagrams are analytically computed at tree-level, even if it is known that they do not lead to the largest signal in the sky. A large hierarchy of masses and couplings are considered for analytical purposes in the most complicated situations, not probing intermediate regimes. Similarly, it is often the case that only particular kinematic limits can be computed analytically, like equilateral or soft configurations, not covering mildly soft ones where the signal-to-noise ratio can be the largest. Ultimately, the rare closed-form expressions in the entirety of the kinematic configurations and for certain diagrams are difficult to exploit in practice because they are written in terms of slowly-converging series. It is worth stressing that there are noticeable exceptions in the literature to each of the points above separately, see~\cite{Pinol:2023oux} for references.\footnote{For instance, signals from strongly-mixed theories at tree level were investigated in e.g.~\cite{Cremonini:2010ua, An:2017hlx, Iyer:2017qzw}, and processes involving multiple particle exchanges can be found in e.g.~\cite{Chen:2009zp, Noumi:2012vr, Aoki:2024uyi}.} Nevertheless, it is fair to say that state-of-the-art technics face limitations, that current predictions do not cover the entire space of cosmological correlators, and that it will always be the fate of analytical methods. A complete dictionary between theory and observables is not yet complete, which can lead to a biased interpretation of current and future cosmological data. With \textsf{CosmoFlow}, our aim is to fill this gap.

\paragraph{What does \textsf{CosmoFlow} compute?} In its current version, the numerical implementation of the cosmological flow takes as an input a theory up to cubic order in the fields, i.e.~a Lagrangian or equivalently a Hamiltonian. This theory can feature an arbitrary number of scalar degrees of freedom with arbitrary dispersion relations, masses and quadratic mixings (or more generally any quadratic theory), coupled through any type of interactions that can have any time dependence. This encompasses all known theories conveniently formulated at the level of fluctuations. The \textsf{CosmoFlow} outputs are \textit{all} tree-level two- and three-point correlators in phase space, i.e.~including those that involve multiple field species and conjugate momenta, at any time during inflation and in all kinematic configurations in Fourier space.

\begin{figure}[h!]
   \centering
   \hspace*{-4cm}
   \includegraphics[width=1\textwidth]{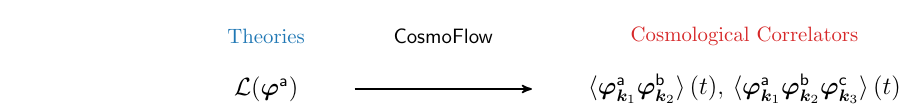}
\end{figure}

\paragraph{Who would benefit from \textsf{CosmoFlow}?} The utility of our code extends to a broad audience. First, it holds potential benefits for theorists interested in formal aspects of cosmological correlators. We have made an effort to render \textsf{CosmoFlow} easy to use. Its elementary implementation enables one to navigate through the modules without effort, and to quickly master their usage, even for those without any experience with Python. \textsf{CosmoFlow} enables one to obtain exact (numerical) results for correlators for which analytical solutions are completely out of reach. Second, the code can be of great interest to primordial cosmologists, from those interested in the phenomenology of primordial non-Gaussianities to model builders of inflationary theories. With this code, we have automatised the process of going from a theory to observables. This enables cosmologists to study inflationary theories and their observational signatures in a systematic fashion and in full generality, without being limited by technicalities. Moreover, the cosmological flow approach allows for a direct insight into the bulk dynamics of correlators, which enables one to elucidate, e.g.~, characteristic time scales to guide physical intuition. Finally, \textsf{CosmoFlow} can be used by cosmologists interested in the later-time Universe as it essentially generates the inputs to be given to pipelines that propagate fluctuations during the Hot Big Bang Universe until the present day, be it correlators, directly the full map-level fields, or simulations.

\vskip 4pt
The outline of the paper is as follows: In Sec~\ref{sec: Short Review of the Cosmological Flow}, we review the minimal (yet necessary) theoretical background for the cosmological flow numerical implementation. The reader familiar with this approach can skip this section. In Sec~\ref{sec: CosmoFlow - Structure and User-guide}, we enter into the details of \textsf{CosmoFlow} exposing how it is organised and constructed. Of interest to the potential \textsf{CosmoFlow} users, we present a detailed user-guide following step by step a simple illustrative example. For aficionados, in Sec~\ref{sec: Applications}, we illustrate the range of possibilities of \textsf{CosmoFlow} with a series of applications accompanied with ready-to-use notebooks. In Sec.~\ref{sec: Performances}, we evaluate the performances of \textsf{CosmoFlow}. Our conclusions are summarised in Sec.~\ref{sec: Conclusion}. 

\section{Short Review of the Cosmological Flow}
\label{sec: Short Review of the Cosmological Flow}

We first begin by reviewing the cosmological flow approach to compute inflationary correlators, setting up the necessary equations and notations for its numerical implementation. 

\subsection{Differential Equations in Time}

At tree-level in perturbative Schwinger-Keldysh diagrammatics, cosmological correlators are computed by solving a closed set of differential equations in time governing their evolution through the entire spacetime. The form of these equations is completely universal and results from first-principle perturbation theory. More details can be found in~\cite{Werth:2023pfl, Pinol:2023oux}.

\paragraph{Equal-time correlators.} Let us consider a set of bulk scalar fields $\bm{\varphi}^\alpha(\bm{x}, t)$ where $\alpha$ runs from 1 to the number of fields $N$, accompanied with their corresponding conjugate momenta $\bm{p}^\beta(\bm{x}, t)$. The theories we consider are directly formulated at the level of inflationary fluctuations. As such, the fields we consider have zero vacuum expectation values and are assumed to be small so that a perturbative approach is well-defined. We assume that all the fields are dynamical degrees of freedom, i.e.~there is no constrained variable nor gauge redundancy. For later convenience, we gather these operators in a phase-space vector $\bm{X}^a \equiv (\bm{\varphi}^\alpha, \bm{p}^\beta)$. Accordingly, Greek indices label \textit{field or momentum} variables (from 1 to $N$) whereas Latin indices label \textit{phase-space} variables (from 1 to $2N$) organised so that a block of field labels is followed by a block of momentum labels, in the same order. We now Fourier transform the spatial coordinates but not the temporal coordinates, defining the operators in Fourier space with \textsf{sans serif} indices such that
\begin{equation}
\bm{X}^a(\bm{x}, t) \equiv \int \frac{\d^3\bm{k}}{(2\pi)^3}\, \bm{X}^{\sf{a}}(\bm{k}, t) \, e^{-i\bm{k}\cdot\bm{x}}\,.
\end{equation}
To avoid clutter in what follows, we will not explicitly write the 3-momentum associated with its operator in Fourier space. Being quantum operators, the phase-space variables obey canonical commutation relations that can be written in the following compact form
\begin{equation}
    [\bm{X}^{\sf{a}}, \bm{X}^{\sf{b}}] = \epsilon^{\sf{ab}}\,,
\end{equation}
where the tensor $\epsilon^{\sf{ab}}$ is defined by $\epsilon^{\sf{ab}} \equiv (2\pi)^3 \delta^{(3)}(\bm{k}_a + \bm{k}_b) \epsilon^{ab}$ where $\epsilon^{ab}$ is the standard antisymmetric bilinear form
\begin{equation}
    \epsilon^{ab} = 
    \begin{pmatrix}
    \bm{0}  & \bm{1} \\
    \bm{-1} & \bm{0}
    \end{pmatrix}\,.
\end{equation}
This matrix defines the real symplectic group $\text{Sp}(2N, \mathbb{R}) \equiv \{M \in \mathcal{M}_{2N\times2N}(\mathbb{R}) \,|\, ^{\text{t}}M \epsilon M = \epsilon\}$ encompassing the symmetries of canonical variables that preserve the canonical commutation relation. As such, the time evolution of the operators $\bm{X}^{\sf{a}}$ is equivalent to an action of the real symplectic group on the phase space. Fundamentally, the cosmological flow can therefore be viewed as the Hamiltonian vector flow in phase space directly implemented at the level of correlators. We are interested in computing correlators of composite operators $\mathcal{O}(\bm{X}^{\sf{a}})$ evaluated at the same time $t$, $\braket{\Omega | \mathcal{O}(\bm{X}^{\sf{a}}) | \Omega}$, where $\ket{\Omega}$ is the vacuum of the full interacting theory whose details are unimportant at this stage. In the following, we will mainly be interested in two- and three-point correlators: $\braket{\bm{X}^{\sf{a}} \bm{X}^{\sf{b}}}$ and $\braket{\bm{X}^{\sf{a}} \bm{X}^{\sf{b}} \bm{X}^{\sf{c}}}$, respectively, where from now on we omit the explicit reference to the vacuum.

\paragraph{Flow equations.} Let us now consider that the operators $\bm{X}^{\sf{a}}$ are described by a Hamiltonian, functional of the phase-space variables. In Fourier space, its most general form can be written as a series expansion in powers of the operators
\begin{equation}
    H = \frac{1}{2!} H_{\sf{ab}} \bm{X}^{\sf{a}} \bm{X}^{\sf{b}} + \frac{1}{3!} H_{\sf{abc}} \bm{X}^{\sf{a}} \bm{X}^{\sf{b}} \bm{X}^{\sf{c}} + \ldots \,,
\end{equation}
where we have adopted the Fourier summation convention stating that repeated indices indicate a sum including
integrals over Fourier modes, i.e.~a Fourier index contraction reads
\begin{equation}
\label{eq: Fourier index contraction}
    A_{\sf{a}} B^{\sf{a}} \equiv \sum_a \int \frac{\d^3 \bm{k}_a}{(2\pi)^3}\, A_a(\bm{k}_a) B^a(\bm{k}_a)\,.
\end{equation}
This notation generalises to multiple indices. Without loss of generality, the tensors $H_{\sf{ab}}, H_{\sf{abc}}, \ldots$ are taken to be symmetric under the exchange of any indices, and their tensor elements can be arbitrary functions of time and momenta. This follows from the fact that \textit{in real space}, the operators $\bm{X}^a$ are Hermitian and the Hamiltonian elements $H_{ab}, H_{abc}, \ldots$ are real. In practice, this amounts to assume a \textit{unitary} time evolution, i.e.~we do not take into account dissipative effects. It is important to highlight that this form of the Hamiltonian is completely universal and captures all known unitary effective field theories for bulk scalar degrees of freedom. From first principles at tree level, one can derive a closed system of differential equations in time satisfied by the correlators. This cosmological flow for the two- and three-point correlation functions reads
\begin{subequations}
\begin{align}
    \frac{\d}{\d t} \langle \bm{X}^{\sf{a}} \bm{X}^{\sf{b}} \rangle &= \tensor{u}{^{\sf{a}}}{_{\sf{c}}} \langle \bm{X}^{\sf{c}} \bm{X}^{\sf{b}} \rangle + \tensor{u}{^{\sf{b}}}{_{\sf{c}}} \langle \bm{X}^{\sf{a}} \bm{X}^{\sf{c}} \rangle\,,\label{eq: 2pt time evolution}\\
    \begin{split}
    \frac{\d}{\d t} \langle \bm{X}^{\sf{a}} \bm{X}^{\sf{b}}\bm{X}^{\sf{c}} \rangle &= \tensor{u}{^{\sf{a}}}{_{\sf{d}}} \langle \bm{X}^{\sf{d}} \bm{X}^{\sf{b}}\bm{X}^{\sf{c}} \rangle
    + \tensor{u}{^{\sf{a}}}{_{\sf{de}}}\langle \bm{X}^{\sf{b}} \bm{X}^{\sf{d}} \rangle\langle \bm{X}^{\sf{c}} \bm{X}^{\sf{e}} \rangle + (2\text{ perms})\,,\label{eq: 3pt time evolution}
    \end{split}
\end{align}
\end{subequations}
where the introduced $u$-tensors are defined by $\tensor{u}{^{\sf{a}}}{_{\sf{b}}} \equiv \epsilon^{\sf{ac}}H_{\sf{cb}}$ and $\tensor{u}{^{\sf{a}}}{_{\sf{bc}}} \equiv \epsilon^{\sf{ad}} H_{\sf{dbc}}$. Eq.~(\ref{eq: 2pt time evolution}) encodes both the full quadratic evolution of $\langle \bm{X}^{\sf{a}}\bm{X}^{\sf{b}} \rangle$ and the quantum properties of the phase-space operators. It couples all two-point correlators, including those which contain conjugate momenta, and accurately encodes all physical effects arising from quadratic operators in the theory. Eq.~(\ref{eq: 3pt time evolution}) is linear and allows the flow of each kinematic configuration to be solved independently. This is a direct consequence of working at tree-level in perturbation theory. These equations are mainly an implementation of the Ehrenfest theorem for correlators and reflect their time evolution in the ``correlator phase space".

\subsection{From Theories to Correlators}
\label{subsec: From Theories to Correlators}

Cosmological correlators satisfy universal flow equations that can be derived by only assuming a unitary time evolution at tree-level in perturbation theory. Yet, the theory dependence enters the specific (time dependent) coefficients that govern these equations, i.e.~the $u$-tensors. We now review how to feed these coefficients given a theory. 

\paragraph{Theories.} Up to cubic order in the fields and in any FLRW spacetime, the most general action in Fourier space takes the following form
\begin{equation}
\begin{aligned}
    S = \frac{1}{2} \int \mathrm{d}t\, a^3 &\left(\bar{\Delta}_{\upalpha\upbeta}\dot{\bm{\varphi}}^{\upalpha}\dot{\bm{\varphi}}^{\upbeta} + \bar{M}_{\upalpha\upbeta}\bm{\varphi}^{\upalpha}\bm{\varphi}^{\upbeta} + 2\bar{I}_{\upalpha\upbeta}\bm{\varphi}^{\upalpha}\dot{\bm{\varphi}}^{\upbeta} \right.\\
    &\left. + \bar{A}_{\upalpha\upbeta\upgamma}\bm{\varphi}^{\upalpha}\bm{\varphi}^{\upbeta}\bm{\varphi}^{\upgamma} + \bar{B}_{\upalpha\upbeta\upgamma}\bm{\varphi}^{\upalpha}\bm{\varphi}^{\upbeta}\dot{\bm{\varphi}}^{\upgamma} + \bar{C}_{\upalpha\upbeta\upgamma}\dot{\bm{\varphi}}^{\upalpha}\dot{\bm{\varphi}}^{\upbeta}\bm{\varphi}^{\upgamma} + \bar{D}_{\upalpha\upbeta\upgamma}\dot{\bm{\varphi}}^{\upalpha}\dot{\bm{\varphi}}^{\upbeta}\dot{\bm{\varphi}}^{\upgamma}  \right)\,,
\end{aligned}
\end{equation}
where $a(t)$ is the scale factor whose time evolution is arbitrary. We recall that any \textsf{sans serif} index contraction contains an integral over Fourier modes, see Eq.~(\ref{eq: Fourier index contraction}). The introduced tensors in general carry arbitrary
momentum and time dependencies, and are assumed to satisfy the following symmetry properties: $\bar{\Delta}_{\upalpha\upbeta} = \bar{\Delta}_{(\upalpha\upbeta)}, \bar{M}_{\upalpha\upbeta} = \bar{M}_{(\upalpha\upbeta)}, \bar{A}_{\upalpha\upbeta\upgamma} = \bar{A}_{(\upalpha\upbeta\upgamma)}, \bar{B}_{\upalpha\upbeta\upgamma} = \bar{B}_{(\upalpha\upbeta)\upgamma}, \bar{C}_{\upalpha\upbeta\upgamma} = \bar{C}_{(\upalpha\upbeta)\upgamma} ,$ and $\bar{D}_{\upalpha\upbeta\upgamma} = \bar{D}_{(\upalpha\upbeta\upgamma)}$. Upon deriving the (rescaled) conjugate momenta to the fields $\bm{p}_{\upalpha}(t) = \tfrac{1}{a^3}\tfrac{\delta S}{\delta \dot{\bm{\varphi}}^{\upalpha}(t)}$ and performing a Legendre transform, the Hamiltonian reads
\begin{equation}
\label{eq: Hamiltonian}
\begin{aligned}
    H = \frac{1}{2}\int \mathrm{d}t\, a^3 &\left( \Delta_{\upalpha\upbeta} \bm{p}^{\upalpha}\bm{p}^{\upbeta} - M_{\upalpha\upbeta}\bm{\varphi}^\upalpha\bm{\varphi}^\upbeta - 2I_{\upalpha\upbeta}\bm{\varphi}^{\upalpha}\bm{p}^{\upbeta} \right.\\
    &\left. - A_{\upalpha\upbeta\upgamma}\bm{\varphi}^{\upalpha}\bm{\varphi}^{\upbeta}\bm{\varphi}^{\upgamma} - B_{\upalpha\upbeta\upgamma}\bm{\varphi}^{\upalpha}\bm{\varphi}^{\upbeta}\bm{p}^{\upgamma} - C_{\upalpha\upbeta\upgamma}\bm{p}^{\upalpha}\bm{p}^{\upbeta}\bm{\varphi}^{\upgamma} - D_{\upalpha\upbeta\upgamma}\bm{p}^{\upalpha}\bm{p}^{\upbeta}\bm{p}^{\upgamma}  \right)\,,
\end{aligned}
\end{equation}
where the relation between bar and unbar tensors can be explicitly found in~\cite{Pinol:2023oux}. The unbar tensors obey the same symmetry properties as the bar ones. It is important to highlight that non-linear contributions to the momentum cancel out in the cubic Hamiltonian, so that deriving the Hamiltonian given a Lagrangian is essentially straightforward. We will illustrate this Legendre transform procedure with concrete examples later on. Given the Hamiltonian~(\ref{eq: Hamiltonian}), the $u$-tensors take the following form
\begin{equation}
    \tensor{u}{^{\sf{a}}}{_{\sf{b}}} = 
    \begin{pmatrix}
    \tensor{I}{_{\upbeta}}{^{\bar{\upalpha}}} & \tensor{\Delta}{^{\bar{\upalpha}}}{_\upbeta} \\
    \tensor{M}{^{\bar{\upalpha}}}{_\upbeta} & \tensor{I}{^{\bar{\upalpha}}}{_\upbeta} - 3H\delta^{\bar{\upalpha}}_{\upbeta}
    \end{pmatrix}\,, 
    \quad 
    \tensor{u}{^{\sf{a}}}{_{\sf{bc}}}
    =
    \begin{Bmatrix}
    \begin{pmatrix}
    -\tensor{B}{_{\upgamma\upbeta}}{^{\bar{\upalpha}}} & -\tensor{C}{^{\bar{\upalpha}}}{_{\upgamma\upbeta}}\\
    3\tensor{A}{^{\bar{\upalpha}}}{_{\upbeta\upgamma}} & \tensor{B}{^{\bar{\upalpha}}}{_{\upgamma\upbeta}}
    \end{pmatrix}\\
    \begin{pmatrix}
    -\tensor{C}{^{\bar{\upalpha}}}{_{\upgamma\upbeta}} & 3\tensor{D}{^{\bar{\upalpha}}}{_{\upbeta\upgamma}}\\
    \tensor{B}{^{\bar{\upalpha}}}{_{\upbeta\upgamma}} & \tensor{C}{_{\upgamma\upbeta}}{^{\bar{\upalpha}}}
    \end{pmatrix}
    \end{Bmatrix}\,,
\end{equation}
where bar indices indicate that the sign of the corresponding momentum $\k$ has been reversed, i.e.~$T^{\bar{\upalpha}} = T^{\upalpha}(-\k_\alpha)$. The factor $- 3H\delta^{\bar{\upalpha}}_{\upbeta}$ comes from the momentum rescaling. In the $\tensor{u}{^{\sf{a}}}{_{\sf{bc}}}$ tensor, the index $\sf{c}$ labels the first (resp. second) matrix within the braces $\{\dots\}$ if it is a field (resp. a momentum) index, and similarly $\sf{a}$ (resp. $\sf{b}$) labels the row (resp. column) in each $2\times2$ block matrix. The particular arrangement of barred indices enables us to extract an overall momentum-conserving delta function. We therefore define $\bm{k}$-dependent tensor coefficients $\tensor{u}{^{a}}{_{b}}$ and $\tensor{u}{^{a}}{_{bc}}$ in the following way
\begin{equation}
    \begin{aligned}
    \tensor{u}{^{\mathsf{a}}}{_{\mathsf{b}}} &= (2\pi)^3\delta^{(3)}(\bm{k}_a - \bm{k}_b)\,\tensor{u}{^{a}}{_{b}}(k)\,,\\
    \tensor{u}{^{\mathsf{a}}}{_{\mathsf{bc}}} &= (2\pi)^3\delta^{(3)}(\bm{k}_a - \bm{k}_b - \bm{k}_c)\,\tensor{u}{^{a}}{_{bc}}(\bm{k}_a, \bm{k}_b, \bm{k}_c)\,.
    \end{aligned}
\end{equation}
where $k = |\bm{k}_a| = |\bm{k}_b|$. Importantly,  in expressions with explicit momentum dependence, the
index labelling the momentum is associated with the corresponding tensor index. Exchanging two indices requires exchanging the corresponding momenta too. 

\paragraph{Explicit flow equations.} We are now ready to state explicit flow equations that can be directly hard coded. By statistical isotropy and homogeneity, the two- and three-point correlators can be written
\begin{equation}
    \begin{aligned}
    \langle \bm{X}^{\sf{a}}\bm{X}^{\sf{b}}\rangle 
    &= (2\pi)^3\delta^{(3)}(\bm{k}_a + \bm{k}_b)\,\Sigma^{ab}(k)\,, \\
    \langle \bm{X}^{\sf{a}} \bm{X}^{\sf{b}} \bm{X}^{\sf{c}}\rangle 
    &= (2\pi)^3\delta^{(3)}(\bm{k}_a + \bm{k}_b + \bm{k}_c)\, B^{abc}(k_a, k_b, k_c)\,.
    \end{aligned}
\end{equation}
The time dependence of the quantities $\Sigma^{ab}$ and $B^{abc}$ is implicit. The three-point function $B^{abc}(k_a, k_b, k_c)$ is real, and $\Sigma^{ab}$ is complex so that it can be split into a real and imaginary parts
$\Sigma^{ab} = \Sigma_{\Re}^{ab} + i \Sigma_{\Im}^{ab}$ that satisfy the following symmetry properties: $\Sigma_{\Re}^{ab} = \Sigma_{\Re}^{ba}$ and $\Sigma_{\Im}^{ab} =- \Sigma_{\Im}^{ba}$. The flow equations for the two-point correlators~(\ref{eq: 2pt time evolution}) being linear and the tensor $\tensor{u}{^{a}}{_{b}}$ having real elements, the flow for $\Sigma_{\Re}^{ab}$ and $\Sigma_{\Im}^{ab}$ can be solved independently
\begin{equation}
\label{eq: explicit 2pt time evolution}
    \frac{\d}{\d t}\Sigma_{\Re, \Im}^{ab}(k) = \tensor{u}{^a}{_c}(k)\, \Sigma_{\Re, \Im}^{cb}(k) + \tensor{u}{^b}{_c}(k)\, \Sigma_{\Re, \Im}^{ac}(k)\,.
\end{equation}
Keeping track of both $\Sigma_{\Re}^{ab}$ and $\Sigma_{\Im}^{ab}$, as the flow equations for the three-point correlators~(\ref{eq: 3pt time evolution}) are sourced by a non-linear term and the tensor $\tensor{u}{^{a}}{_{bc}}$ having real elements, we obtain
\begin{gather}
\label{eq: explicit 3pt time evolution}
    \begin{aligned}
    \frac{\d}{\d t} B^{abc}(k_a, k_b, k_c) &= \tensor{u}{^a}{_d}(k_a)B^{dbc}(k_a, k_b, k_c) + \tensor{u}{^b}{_d}(k_b)B^{adc}(k_a, k_b, k_c) + \tensor{u}{^c}{_d}(k_c)B^{abd}(k_a, k_b, k_c) \\
    &+ \tensor{u}{^a}{_{de}}(\bm{k}_a, \bm{k}_b, \bm{k}_c)\Sigma_{\Re}^{db}(k_b)\Sigma_{\Re}^{ec}(k_c) - \tensor{u}{^a}{_{de}}(\bm{k}_a, \bm{k}_b, \bm{k}_c)\Sigma_{\Im}^{db}(k_b)\Sigma_{\Im}^{ec}(k_c) \\
    &+ \tensor{u}{^b}{_{de}}(\bm{k}_b, \bm{k}_a, \bm{k}_c)\Sigma_{\Re}^{ad}(k_a)\Sigma_{\Re}^{ec}(k_c) - \tensor{u}{^b}{_{de}}(\bm{k}_b, \bm{k}_a, \bm{k}_c)\Sigma_{\Im}^{ad}(k_a)\Sigma_{\Im}^{ec}(k_c) \\
    &+ \tensor{u}{^c}{_{de}}(\bm{k}_c, \bm{k}_a, \bm{k}_b)\Sigma_{\Re}^{ad}(k_a)\Sigma_{\Re}^{be}(k_b) - \tensor{u}{^c}{_{de}}(\bm{k}_c, \bm{k}_a, \bm{k}_b)\Sigma_{\Im}^{ad}(k_a)\Sigma_{\Im}^{be}(k_b)\,,
    \end{aligned}
\raisetag{35pt}
\end{gather}
where we have deliberately written all the permutations to avoid ambiguity in the index ordering. From the form of Eq.~(\ref{eq: explicit 3pt time evolution}), it is clear that $B^{abc}$ is fully symmetric. Therefore, the total number of independent equations to solve up to three-point correlators are\footnote{It should be noted that $\Sigma^{ab}_\Im$ is completely fixed by the commutation relation so that in principle its flow equation does not need to be solved.}
\begin{equation}
    \underbrace{\frac{2N(2N+1)}{2}}_{\Sigma^{ab}_\Re} + \underbrace{\frac{2N(2N-1)}{2}}_{\Sigma^{ab}_\Im} + \underbrace{\frac{2N(2N+1)(2N+2)}{3!}}_{B^{abc}} = \mathcal{O}((2N)^3)\,.
\end{equation}
The aim of \textsf{CosmoFlow} is to systematically solve the flow equations (\ref{eq: explicit 2pt time evolution}) and~(\ref{eq: explicit 3pt time evolution}).

\paragraph{Initial conditions.} For the system to be complete, we need initial conditions for $\Sigma_{\Re}^{ab}, \Sigma_{\Im}^{ab}$ and $B^{abc}$. In the far past, namely when all modes of interest are well inside the horizon, correlators can be initialised analytically.\footnote{Note that the time evolution given by the flow equations is valid in all FLRW spacetimes, but unambiguously initialising the correlators requires a shrinking comoving Hubble radius $|a H|^{-1}$.} Essentially, the quadratic theory is dominated by spatial gradients, mass terms can be neglected, and mixing interactions become irrelevant so that the system decouples and approaches that of a set uncoupled massless degrees of freedom, i.e.~it asymptotically reaches the vacuum state. Initial conditions can therefore be readily derived analytically and are completely fixed by the tensors appearing in~(\ref{eq: Hamiltonian}), see~\cite{Pinol:2023oux} for explicit expressions considering the Bunch-Davies state. Naturally, any other state can also be implemented. As such, deriving these tensors fully determines the theory, hence the flow equations, \textit{and} the initial conditions. 

\paragraph{$i\epsilon$ prescription.} The standard approach of computing cosmological correlators requires deforming the time integration contour to the imaginary direction at early times, with fields appearing in the integral defined by analytical continuation $t \rightarrow t(1\pm i\epsilon)$, to achieve UV convergence. This procedure maps the vacuum of the free theory to the full theory vacuum in the infinite past and is at the core of the in-in formula. The cosmological flow approach completely bypasses this complex deformation of the integrand as the correlators are evolved on the \textit{real time axis}. The $i\epsilon$ prescription is transferred to the derivation of analytical initial conditions. However, an alternative elementary approach is to realise that the $i\epsilon$ defines the asymptotic vacuum state in which all interactions are turned off. Therefore, for the numerical implementation of the cosmological flow, we also propose a \textit{numerical} $i\epsilon$ prescription that consists in initialising the three-point correlators to zero and adiabatically switching on interactions, be they cubic interactions or quadratic mixings, allowing various correlators to build at early times before reaching the attractor solution given by the flow equations. 

\section{\textsf{CosmoFlow}: Structure and User-guide}
\label{sec: CosmoFlow - Structure and User-guide}

Built upon the cosmological flow approach, \textsf{CosmoFlow} automatically computes cosmological correlators. Given a certain theory, i.e.~knowledge of the tensors $\tensor{u}{^{\sf{a}}}{_{\sf{b}}}$ and $\tensor{u}{^{\sf{a}}}{_{\sf{bc}}}$, it is straightforward to implement a routine that numerically solves the flow equations. In this section, we show in details how \textsf{CosmoFlow} has been implemented, and present a user-guide for those seeking to initiate the use of the code.

\subsection{Prerequisites}

Aiming for simplicity above all, \textsf{CosmoFlow} has minimal prerequisites.

\paragraph{Python.} The code \textsf{CosmoFlow} needs a working Python installation. During the course of the development, we used Python 3 (which we recommend) although it should run without problem on Python 2.7 (or below). For convenience, we recommend to use a complete Python distribution such as \href{https://www.anaconda.com/download}{\textsf{Anaconda}}, which comes with all the packages used by the code. It also contains a nice interactive development environment and includes \href{https://jupyter.org/}{\textsf{Jupyter}} that enables to run all the \textsf{CosmoFlow} notebooks. The development of \textsf{CosmoFlow} was performed on a Macbook Pro with M1 CPU and running MacOS 11.6 Big Sur, and the code is available for Linux, MacOS and Windows systems.

\paragraph{Packages.} We have precisely designed \textsf{CosmoFlow} so that it only uses elementary and commonly-used Python packages, as well as employs no sophisticated Python implementation nor hard coding. The Python packages used in \textsf{CosmoFlow} and/or in the numerous notebooks accompanying the code are \href{https://numpy.org/}{\textsf{Numpy}}~\cite{harris2020array}, \href{https://matplotlib.org/}{\textsf{Matplotlib}}~\cite{Hunter:2007}, and \href{https://scipy.org/}{\textsf{Scipy}}~\cite{2020SciPy-NMeth}. The modules \href{https://docs.python.org/3/library/time.html}{\textsf{time}} and \href{https://tqdm.github.io/}{\textsf{tqdm}} are also used in tutorial notebooks only, and are for timing the computation and displaying a progress bar, respectively. These modules are not essential and their use can be avoided by commenting out the appropriate code lines. All these packages can be installed using \href{https://pip.pypa.io/en/stable/}{\textsf{pip}}. If you use \href{https://docs.conda.io/en/latest/}{\textsf{Conda}} or similar Python environments, they usually come with their own package manager to install packages. Have in mind that the web is a goldmine of useful resources to help installing and using these packages.

\subsection{Installation}

\textsf{CosmoFlow} is distributed via its \href{https://github.com/deniswerth/CosmoFlow/tree/main}{GitHub public repository}. The easy way to import the code on your machine is to click the green button ``code" and then ``download ZIP". The zipped folder contains the entire repository. The code can also be downloaded as a complete git repository with
\begin{lstlisting}[frame={}]
    $ git clone https://github.com/deniswerth/CosmoFlow.git CosmoFlow
\end{lstlisting}
The repository also contains the current paper source code, many code duplicates where various theories have been pre-implemented, as well as numerous \href{https://jupyter.org/}{\textsf{Jupyter}} notebooks, see Sec.~\ref{sec: Applications}.

\subsection{Code Architecture}

We have designed \textsf{CosmoFlow} to be flexible, in the sense that it can be easily modified, improved and generalised. We believe that this goal can be reached by having a simple architecture and no hard coding. The \textsf{CosmoFlow} implementation has a clear structure, separated in modules that fully exploit the object-oriented programming paradigm of Python (coded as \href{https://docs.python.org/3/tutorial/classes.html}{Python classes}) with well defined tasks:

\begin{itemize}
    \item \textbf{Parameters.py} contains the \textsf{parameters} class that defines all the necessary variables and arrays, including the time grid array in $e$-folds \textsf{N\_load} on which the time-dependent parameters are defined, and parameters entering the theory, e.g. Hubble parameter and coupling constants. This class defines interpolated continuous functions out of the given inputs using the function \textsf{interp1d} from \textsf{Scipy}. It enables to treat both analytical or numerical (given by an array evaluated on a discrete set of points) time-dependent parameters. It is important to define all parameters in this class, even the constant ones, as some of them (typically coupling constants) will be switched on adiabatically at early times to implement the numerical $i\epsilon$ prescription. For every argument of this class, say \textsf{variable}, one should: (i) define it as an argument of the class \textsf{variable\_load}, (ii) state it as a variable within the class using \textsf{self.variable\_load}, (iii) create a continuous function \textsf{self.variable\_f}, and (iv) add \textsf{variable\_f} to the \textsf{interpolated} list that contains all interpolated continuous functions, in the same order as given by the arguments of the class.
    
    \item \textbf{Theory.py} contains the \textsf{theory} class that defines the $u$-tensors given the tensors appearing explicitly in the Hamiltonian~(\ref{eq: Hamiltonian}). All tensor elements of this class are continuous functions that can be evaluated at a given time. This class will be called at each integration step of the flow equations. \textsf{Theory} takes as inputs \textsf{N} (time in $e$-folds at which all the parameters are defined), \textsf{Nfield} (number of fields), and \textsf{interpolated} (list of all the interpolated functions given by \textsf{parameters}). 

    \vskip 4pt
    For each continuous function, say \textsf{variable\_f}, one needs to define a variable within the class \textsf{self.variable} and a function \textsf{variable\_f} of \textsf{N} using the list \textsf{interpolated}. It is important to keep the same continuous parameter indexing as defined in \textsf{Parameters}. The quadratic-theory tensors $\Delta_{\upalpha\upbeta}, M_{\upalpha\upbeta}$ and $I_{\upalpha\upbeta}$ and the cubic-theory tensors $A_{\upalpha\upbeta\upgamma}, B_{\upalpha\upbeta\upgamma}, C_{\upalpha\upbeta\upgamma}$ and $D_{\upalpha\upbeta\upgamma}$ must be fed according to the considered theory. Recall that Python indexing starts with $0$ and that these tensors must satisfy specific symmetry properties that should be well implemented, see Sec.~\ref{subsec: From Theories to Correlators}. Defining the tensors $\tensor{u}{^{\mathsf{a}}}{_{\mathsf{b}}}$ and $\tensor{u}{^{\mathsf{a}}}{_{\mathsf{bc}}}$ is done automatically.
    
    \item \textbf{Solver.py} contains the \textsf{solver} class that defines the initial conditions for the two- and three-point correlators, and the flow equations. Being universal regardless the theory, these functions are hard-coded. This class takes as inputs \textsf{Nspan} (time grid array in $e$-folds on which the flow equations are solved), \textsf{Nfield} (number of fields), \textsf{interpolated} (list of all the interpolated functions given by \textsf{parameters}), \textsf{Rtol} (relative tolerance for the numerical integrator), and \textsf{Atol} (absolute tolerance for the numerical integrator). The function \textsf{f\_solution} contains the \textsf{solve\_ivp} solver to integrate the flow equations. Various integrators can be used as a \textsf{method} option, see the \href{https://docs.scipy.org/doc/scipy/reference/generated/scipy.integrate.solve_ivp.html}{\textsf{solve\_ivp} documentation} for more details. By default, RK45 is used.
\end{itemize}

\begin{figure}[h!]
\begin{center}
\begin{tikzpicture}
			
\node[nearnodes] (T) [model1] {
    {\large\textcolor{pyblue}{\textsf{Theory}}}
    \begin{itemize}[leftmargin=.19in]
        \item $^{\textcolor{pyred}{\ddagger}}$Define the tensors $\Delta_{\upalpha\upbeta}$, $M_{\upalpha\upbeta}$, $I_{\upalpha\upbeta}$, $A_{\upalpha\upbeta\upgamma}$, $B_{\upalpha\upbeta\upgamma}$, $C_{\upalpha\upbeta\upgamma}$ and $D_{\upalpha\upbeta\upgamma}$
	\item Automatically feed the tensors $\tensor{u}{^{\mathsf{a}}}{_{\mathsf{b}}}$ and $\tensor{u}{^{\mathsf{a}}}{_{\mathsf{bc}}}$ whose elements can be continuously evaluated at any time
    \end{itemize}
    };
		
\node[nearnodes] (P) [model1, left = of T] {
    {\large\textcolor{pyblue}{\textsf{Parameters}}}
    \begin{itemize}[leftmargin=.19in]
        \item $^{\textcolor{pyred}{\ddagger}}$Define all time-dependent parameters of the theory
	\item $^{\textcolor{pyred}{\ddagger}}$Create interpolated continuous functions stored in the list \textsf{interpolated}
    \end{itemize}
    };
			
\node[nearnodes] (S) [model1, right = of T] {
    {\large\textcolor{pyblue}{\textsf{Solver}}}
    \begin{itemize}[leftmargin=.19in]
        \item Define initial conditions
        \item Define flow equations
        \item Solve the complete set of flow equations
    \end{itemize}
    };
		
\draw [arrow] (T) -- (P);
\draw [arrow] (S) -- (T);
			
\node[draw, dashed, inner xsep = 2em, inner ysep = 2em, fit = (P) (T) (S)] (box) {};
\node[fill=white] at (box.north) {\textsf{CosmoFlow}};
			
\end{tikzpicture}
\end{center}
\caption{General overview of the \textsf{CosmoFlow} architecture with the main features. The class \textsf{Solver} calls \textsf{Theory} at each integration step, and \textsf{Theory} uses continuous parameters defined in \textsf{Parameters}. Tasks decorated with ${\textcolor{pyred}{\ddagger}}$ are theory dependent and must be implemented manually.}
\label{fig: CosmoFlow Architecture}
\end{figure}

Aiming for flexibility and user-friendliness, physical quantities (number of fields, coupling constants, etc) or numerical parameters (integrator, relative tolerance, etc) that are likely to be modified in different theories or usages are localised in a small number of well-identified places in the code that the user can access easily. A general overview of the \textsf{CosmoFlow} architecture is given in Fig.~\ref{fig: CosmoFlow Architecture}. An explicit and detailed example of module implementation is given in Sec.~\ref{subsec: Basic Usage}. Once these files have been implemented, they do not need to be modified further.\footnote{Although it may seem evident, do not forget to save these files after implementation.} Executing \textsf{CosmoFlow} for various applications is performed in a separate file (Python script or notebook).

\subsection{Basic Usage}
\label{subsec: Basic Usage}

Let us now introduce the basic functionalities of \textsf{CosmoFlow} and lead the user through its first step-by-step implementation of a concrete theory. In this section, we first introduce a toy theory that serves as a simple example of how to use \textsf{CosmoFlow}. We then explicitly show how to implement the modules related to this theory and run the code. A blank template for the \textsf{CosmoFlow} modules to implement given a theory is provided in the \href{https://github.com/deniswerth/CosmoFlow/tree/main/CosmoFlow/BlankTemplate}{Github repository}. This template also contains various function’s docstring, listing input and output parameters.

\vskip 4pt
We consider a simple massless $\dot{\varphi}^3$ theory in de Sitter. The action written in cosmic time is
\begin{equation}
\label{eq: dphi3 action}
    S = \int \d t\d^3x\, a^3\left[-\frac{1}{2}(\partial_\mu \varphi)^2 - \frac{g}{3!}\,\dot{\varphi}^3\right]\,,
\end{equation}
where $g$ is a coupling constant and an overdot indicates a derivative with respect to cosmic (physical) time. Implementing the theory in \textsf{CosmoFlow} requires switching to the Hamiltonian. The rescaled conjugate momentum associated with the field $\varphi$ is
\begin{equation}
\label{eq: massless dphi3 conjugate momentum}
    p_\varphi \equiv \frac{1}{a^3} \frac{\delta S}{\delta \dot{\varphi}} = \dot{\varphi} - \frac{g}{2}\dot{\varphi}^2\,.
\end{equation}
At cubic order in the fields, it is enough to consider the \textit{linear} conjugate momentum. However, for illustration purposes, we explicitly show that non-linear contributions cancel out. Inverting the relation~(\ref{eq: massless dphi3 conjugate momentum}) perturbatively, and performing the Legendre transform gives
\begin{equation}
    \begin{aligned}
    H &= \int \d t \d^3x \,a^3 \left[p_\varphi \dot{\varphi} - \mathcal{L}\right] \\
    &= \int \d t \d^3x \,a^3 \left[p_\varphi^2 + \textcolor{pyred}{\frac{g}{2}\,p_\varphi^3} - \frac{1}{2}\,p_\varphi^2 - \textcolor{pyred}{\frac{g}{2}\,p_\varphi^3} + \frac{1}{2}\frac{(\partial_i \varphi)^2}{a^2} + \frac{g}{3!}\,p_\varphi^3 + \mathcal{O}\left(p_\varphi^4\right) \right] \\
    &= \int \d t \d^3x \,a^3 \left[\frac{1}{2}\,p_\varphi^2 + \frac{1}{2}\frac{(\partial_i \varphi)^2}{a^2} + \frac{g}{3!}\,p_\varphi^3 + \mathcal{O}\left(p_\varphi^4\right) \right]\,.
    \end{aligned}
\end{equation}
In Fourier space, this Hamiltonian can be recasted in the form~(\ref{eq: Hamiltonian}) and the identification of the tensors (in this case containing a single element) yields
\begin{equation}
    \Delta_{\varphi\varphi} = 1 \,,\quad M_{\varphi\varphi} = -\frac{k^2}{a^2}\,,\quad D_{\varphi\varphi\varphi} = -\frac{g}{3}\,.
\end{equation}
We recall that these elements contain momentum-conserving delta functions that should be taken into account in the presence of spatial gradients. Knowledge of these tensors is all that is needed to implement the theory in \textsf{CosmoFlow}. When implementing a new theory (or class of theories), our advice is to create a dedicated folder that contains all the \textsf{CosmoFlow} modules and the additional scripts that execute the code. As an example, a new folder called \textsf{Massless\_dphi3} with the $\dot{\varphi}^3$ theory can be created as follows
\begin{lstlisting}[frame={}]
    $ cd Documents # Enter in the Documents folder (or any other)
    $ mkdir Massless_dphi3 # Create a new directory
    $ cd Massless_dphi3 # Go inside the directory
    $ touch Parameters.py # Create a new .py file named Parameters.py
    $ open Parameters.py # Open the file with default application
\end{lstlisting}
This folder (with the required files) already exists in the GitHub repository. All the parameters of the theory should be specified and defined in the \textsf{parameters} class in the \textsf{Parameters.py} file. In our case, we have the time grid array on which the parameters are defined \textsf{N\_load}, the Hubble parameter \textsf{H\_load}, and the cubic coupling constant \textsf{g\_load}. We use the suffix \textsf{\_load} to indicate that the parameters are imported and not yet interpolated (as opposed to \textsf{\_f} which means that the corresponding variables are continuous functions). The \textsf{Parameters.py} file is the following\footnote{The character \# is used for comments (in \textcolor{pygreen}{green}) so everything following it in a given line should be ignored.}

\begin{lstlisting}
from scipy.interpolate import interp1d # import interp1d from Scipy package

class parameters():
	"""
	This class takes as inputs the parameters of the theory and creates interpolated continuous functions
	"""

	def __init__(self, N_load, H_load, g_load):

		#Importing pre-computed parameters of the theory
		self.N_load = N_load
		self.H_load = H_load
		self.g_load = g_load

		#Creating continuous functions out of the imported parameters
		self.H_f = interp1d(self.N_load, self.H_load, bounds_error = False, kind = 'cubic', fill_value = "extrapolate")
		self.g_f = interp1d(self.N_load, self.g_load, bounds_error = False, kind = 'cubic', fill_value = "extrapolate")

		#Creating interpolating list containing continuous functions
		self.interpolated = [self.H_f, self.g_f]

	def output(self):
		"""
		Returns the interpolated continous functions that can be evaluated at any N (time in e-folds)
		"""
		return self.interpolated
\end{lstlisting}
Once all parameters have been defined, one needs to define the \textsf{theory} class in the \textsf{Theory.py} file. We start by defining the class and all variables with the following lines

\begin{lstlisting}
import numpy as np # import Numpy package for vectorisation
from scipy.misc import derivative # import derivative from Scipy package

class theory():
	"""
	This class defines the theory: (i) time-dependent functions, (ii) tensors Delta, M, I, A, B, C and D, and (iii) tensors u^A_B and u^A_BC
	All functions have been optimised to be computed once at each integration step
	"""

	def __init__(self, N, Nfield, interpolated):
		self.N = N
		self.Nfield = Nfield
		self.interpolated = interpolated # (H_f, g_f)

		#Evaluate the parameters at N
		N = self.N
		self.H = self.H_f(N)
		self.g = self.g_f(N)
\end{lstlisting}
Additional functions that have explicit dependence on \textsf{N} (e.g. the scale factor) or that depend on the previously defined parameters (e.g. derivative of the Hubble rate with respect to cosmic time) can also be defined, see \href{https://github.com/deniswerth/CosmoFlow/blob/main/CosmoFlow/Massless_dphi3/Theory.py}{\faGithub} for more details. Because correlators containing conjugate momenta to the fields decay with different rates on super-horizon scales depending on the number of momenta, we rescale the corresponding correlators using the function \textsf{scale}. This procedure is not necessary but improves numerical stability. In order for the parameters to be evaluated continuously at each integration step, all defined quantities are defined as functions in the following way

\begin{lstlisting}
        #Define continuous functions
	def H_f(self, N):
		return self.interpolated[0](N)

	def g_f(self, N):
		return self.interpolated[1](N)
\end{lstlisting}
It is important to keep the same order for continuous variables encoded in the list \textsf{interpolated}. These functions allow us to feed the quadratic and cubic tensors appearing in the Hamiltonian~(\ref{eq: Hamiltonian}), e.g.

\begin{lstlisting}
	def M_ab(self, k):
		Nfield = self.Nfield
		Mab = np.eye(Nfield)
		Mab[0, 0] = -k**2/(self.a**2)
		return Mab
\end{lstlisting}
Note that if a tensor index exceeds $\textsf{Nfield}$, an error message would pop up while executing the code. Given these tensors, the $u$-tensors are defined automatically. We invite new users to go through these files in the repository while reading these lines at the same time. The \textsf{Solver.py} file containing the numerical integrator should not be modified. The philosophy of \textsf{CosmoFlow} is that once these files have been implemented for a specific theory, the user would never have to modify them again. At this point, we are now ready to execute the code for the first time.

\paragraph{My first run.} We now show how to run the code and pass different parameters to \textsf{CosmoFlow} to obtain the first computed correlator. This is done in a separate Python script or Jupyter notebook. The following Python script (\textsf{MyFirstRun.py} in the associated folder) demonstrates the basic execution of \textsf{CosmoFlow}. 

\begin{lstlisting}
import numpy as np # import Numpy package for vectorisation
import matplotlib.pyplot as plt # import matplotlib for visualisation

# Import CosmoFlow modules
from Parameters import parameters
from Theory import theory
from Solver import solver

# Define the numerical i\epsilon prescription
def adiabatic(N_load, DeltaN):
    return (np.tanh((N_load + DeltaN - 1)/0.1) + 1)/2

n = 10000 # Number of points for the parameter evaluation
N_load = np.linspace(-10, 20, n) # Time grid array in e-folds for the parameters
DeltaN = 4 # Number of e-folds before horizon crossing

# Theory 
g_load = 1 * np.ones(n) * adiabatic(N_load, DeltaN) # Cubic coupling constant
H_load = np.ones(n) # Hubble scale

# Load the parameters and define continuous functions
param = parameters(N_load, H_load, g_load) # Load the class parameters
interpolated = param.output() # Define list with continuous parameters

# Numerical parameters
Nspan = np.linspace(-10, 20, 500) # Time span in e-folds for the numerical integration
Nfield = 1 # Number of fields
Rtol, Atol = 1e-4, 1e-180 # Relative and absolute tolerance of the integrator
N_exit = 0 # Horizon exit for a mode
Ni, Nf = N_exit - DeltaN, 10 # Sets initial and final time for integration
N = np.linspace(Ni, Nf, 1000) # Define the time array for output correlators

# Initialise the integrator
theo = theory(N = Nspan, Nfield = Nfield, interpolated = interpolated)

# Kinematic configuration
k = theo.k_mode(N_exit) # Mode corresponding to N = 0 horizon exit
k1, k2, k3 = k, k, k # Kinematic configuration for 3-pt function (here equilateral)

# Solve flow equations
s = solver(Nspan = N, Nfield = Nfield, interpolated = interpolated, Rtol = Rtol, Atol = Atol)
f = s.f_solution(k1 = k1, k2 = k2, k3 = k3)

# Plot correlators
plt.semilogy(N, np.absolute(f[0][0, 0])) # field-field correlator
plt.semilogy(N, np.absolute(f[6][0, 0, 0])) # field-field-field correlator
plt.show()
\end{lstlisting}
If everything went smoothly, by executing this script using the following lines
\begin{lstlisting}[frame={}]
    $ cd ... # Go to Massless_dphi3 directory
    $ python MyFirstRun.py # Execute the script
\end{lstlisting}
you should see appearing a window with the correlators $\Sigma_\Re^{\varphi\varphi}(k)$ and $B^{\varphi\varphi\varphi}(k, k, k)$ (equilateral kinematic configuration) as function of time, expressed in number of $e$-folds with respect to horizon crossing of the mode $k$. Let us have a look at this script and give some explanations. The first thing to do is to import the three \textsf{CosmoFlow} modules. They should be stored in the same folder as the executing script. Otherwise their location on your machine should be specified. The function \textsf{adiabatic} encodes the numerical $i\epsilon$ prescription in a simple manner: it is a window function starting at zero and ending at unity centered around one $e$-fold after initialising the correlators with width $\sigma_N=0.1$ $e$-folds. The parameter $\textsf{DeltaN}$ is the number of $e$-folds before horizon crossing, and specifies the initial time for the correlators.
Turning on interactions can be considered an adiabatic process when the function \textsf{adiabatic} varies on a timescale much longer than the relevant timescale of the dynamics inside the horizon. Considering a unit sound speed without loss of generality, hence with modes behaving as $e^{i k \tau}$, one should require $1/k \ll (\Delta \tau)^{\textrm{adiabatic}} \simeq 3 \sigma_N/(a H)$, at the central time of the \textsf{adiabatic} function. One thus obtain the criterion $\sigma_N \gg e^{-\Delta N}$. Sharper transitions would excite negative frequency modes and would not correspond to the Bunch-Davies vacuum.
The function \textsf{adiabatic} should multiply all cubic coupling constants. This is the very reason why all parameters are defined as functions on a time grid, even though certain coupling constants are actually taken to be constant, with e.g. \textsf{np.ones(n)}. After defining the parameters of the theory and numerical parameters, these are loaded using \textsf{parameters} and the list \textsf{interpolated} is created. The flow equation integrator is initialised by calling \textsf{theory} which takes \textsf{interpolated} as an argument. Before solving the flow equations, a specific kinematic configuration needs to be chosen. Here, we take an equilateral one $k_1=k_2=k_3 = k$ for simplicity. Detailed examples of different kinematic configurations can be found in Sec.~\ref{subsec: Kinematic Dependence}. We use the convention that the scale factor as a function of the number of $e$-folds is given by $a(N) = e^{N}$, that arbitrarily fixes $a=1$ at $N=0$. This way, the default $k$ is chosen such that the corresponding mode crosses its horizon at \textsf{N\_exit}$=0$, i.e.~$k = a H$. Therefore, in scale-invariant theories, there is a one-to-one correspondence between a mode $k$ and the time $t_\star$ at which it crosses the horizon. The corresponding scale is determined in this way by the use of the function \textsf{k\_mode} in the \textsf{Theory} class. It is important to keep this in mind when varying the kinematic configuration, i.e.~when all modes are not the same. Finally, solving the flow equations is done by calling \textsf{solver} and executing the function \textsf{f\_solution}. 

\vskip 4pt
All correlators are stored in the object \textsf{f} in the following way:
\begin{itemize}
    \item Setting the first index to $0, 1$ or $2$ selects the modes $k_1, k_2$ or $k_3$, respectively, corresponding to the \textit{real part of the two-point correlators}. The next layer of indices selects the correlator, e.g.~\textsf{f[1][0, 1]} would give $\Sigma_\Re^{\varphi p_\varphi}(k_2)$.
    \item Setting the first index to $3, 4$ or $5$ selects the modes $k_1, k_2$ or $k_3$, respectively, corresponding to the \textit{imaginary part of the two-point correlators}. The next layer of indices selects the correlator, e.g.~\textsf{f[3][0, 0]} would give $\Sigma_\Im^{\varphi \varphi}(k_1)$.
    \item Setting the first index to $6$ corresponds to selecting the three-point correlators, e.g.~\textsf{f[6][0, 0, 0]} encodes $B^{\varphi\varphi\varphi}(k_1, k_2, k_3)$. One can explicitly check that this object is fully symmetric. 
\end{itemize}
These objects are single-dimensional arrays of the same size as \textsf{N}. The last element of these arrays, which in most applications is the desired value for the correlator at the end of inflation, are given by e.g.~\textsf{f[0][0, 0, 0][-1]} for $B^{\varphi\varphi\varphi}(k_1, k_2, k_3)$. The rest of the script file is rather straightforward as each line is commented appropriately. 
Eventually, for users interested in only computing two-point correlation functions, entries of cubic tensors can be set to zero (which speeds up computations).

\begin{figure}[h!]
  \centering
  \subfloat{\includegraphics[width=0.45\textwidth]{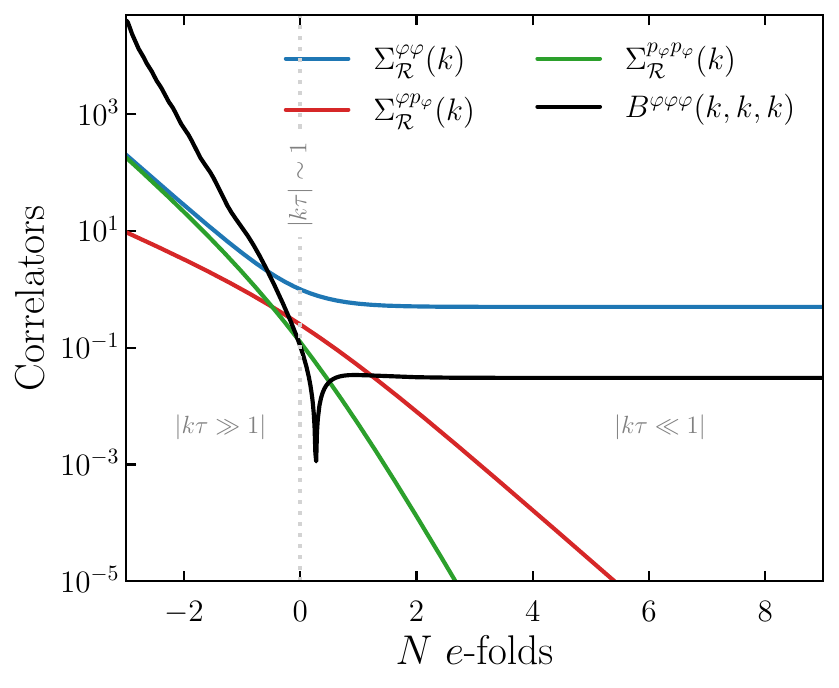}}
  \hspace*{0.2cm}
  \subfloat{\includegraphics[width=0.45\textwidth]{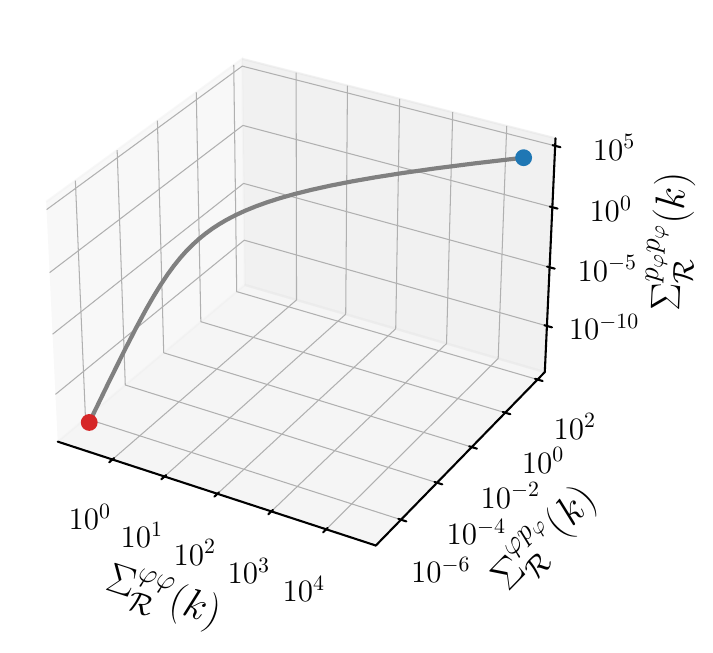}}
  \vspace*{0.2cm}
  \caption{\textit{Left}: Two- and three-point correlation functions as function of number of $e$-folds $N$, expressed so that horizon crossing $|k\tau|\sim 1$ (in \textcolor{lightgray}{gray} dashed line) corresponds to $N=0$. We only show the real part of correlators for simplicity. The \textcolor{pyblue}{blue}, \textcolor{pyred}{red} and \textcolor{pygreen}{green} lines correspond to the real part of $\braket{\varphi_{k} \varphi_{-k}}'$, $\braket{\varphi_{k} p_{\varphi -k}}'$ and $\braket{p_{\varphi k} p_{\varphi -k}}'$, respectively. The black solid line corresponds to $\braket{\varphi_{k_1} \varphi_{k_2} \varphi_{k_3}}'$ in the equilateral configuration $k_1=k_2=k_3$. \textit{Right}: Two-point correlation function flow trajectory in the three-dimensional phase space. The initial point is denoted in \textcolor{pyblue}{blue} and the \textcolor{pyred}{red} point indicates a time sufficiently well ouside the horizon $|k\tau| \ll 1$. Numerical parameters are set to $\Delta N = 4$ and $\Delta_r = 10^{-4}$. [\href{https://github.com/deniswerth/CosmoFlow/blob/main/CosmoFlow/Massless_dphi3/MyFirstRun.ipynb}{\faGithub}]}
  \label{fig: Masslessdphi3 time evolution}
\end{figure}

\vskip 4pt
In Fig.~\ref{fig: Masslessdphi3 time evolution}, we show the time evolution of various correlators, as a result of a single \textsf{CosmoFlow} run. For pedagogical reasons, we also give a ready-to-use Jupyter notebook that can be found by clicking on the following icon \href{https://github.com/deniswerth/CosmoFlow/blob/main/CosmoFlow/Massless_dphi3/MyFirstRun.ipynb}{\faGithub}, or directly in the GitHub repository. This notebook contains executable cells to generate Fig.~\ref{fig: Masslessdphi3 time evolution}. This figure showcases that we can get direct access to the bulk dynamics of various correlators, from initial times deep inside the horizon to later times, typically when correlators of massless fields freeze on super-horizon scales. Scaling laws as function of the scale factor, in $\propto a^p$, can be directly read off, both at high $\omega \gg H$ and low $\omega \ll H$ energies. The transition located around horizon crossing, corresponding to a turn in the phase space, is well visible, and changes the power-law scaling of various correlators. The dominant massless field mode freezes on super-horizon scales $\varphi_k \propto a^0$ whereas the conjugate momentum decays as $p_{\varphi k} \propto a^2$.\footnote{Remember that it has been rescaled in the Hamiltonian and that an extra rescaling has been implemented to improve numerical robustness.} The region well outside the horizon is equivalent to being close to the infinite future surface of de Sitter spacetime as correlators of massless fields freeze. In practical cases, late-time correlators are evaluated at that time.

\paragraph{Parallelisation.} Computing cosmological correlators can be computationally expensive, e.g., when it comes to scan over certain kinematic configurations. However, an important feature of the cosmological flow is that the flow equations for a certain kinematic configuration are independent of one another. In practice, it means that every \textsf{CosmoFlow} run can be performed completely independently of one another, and a fully parallelised implementation is possible. Examples of implementations using parallelisation are given in Secs.~\ref{subsec: Theory Dependence} and~\ref{subsec: Kinematic Dependence}. While developing the code, we have tested parallelisation using the modules \href{https://mpi4py.readthedocs.io/en/stable/}{\textsf{Mpi4Py}} (it requires a working MPI installation such as \href{https://www.open-mpi.org/}{\textsf{openMPI}}) and \href{https://joblib.readthedocs.io/en/stable/}{\textsf{joblib}}. Note that these packages are not needed to run \textsf{CosmoFlow} in general. Nonetheless, this feature makes it possible to run \textsf{CosmoFlow} on multiple cores rather than in serial on a laptop-class hardware. You are welcome to contact the developers for more details.

\subsection{Possible Issues}
\label{subsec: Possible Issues}

At the core of the cosmological flow implementation, some things can go wrong if not done correctly and several tricks are regularly used. It is crucial for the new user to know them. In what follows, we present some common issues while using \textsf{CosmoFlow} and how to avoid them. We encourage the user to report any other issue or bug. The performed tests and generated figures of this section can be found in the corresponding Jupyter notebook by clicking on the following icon: \href{https://github.com/deniswerth/CosmoFlow/blob/main/CosmoFlow/Massless_dphi3/PossibleIssues.ipynb}{\faGithub}.

\begin{figure}[h!]
  \centering
  \includegraphics[width=0.6\textwidth]{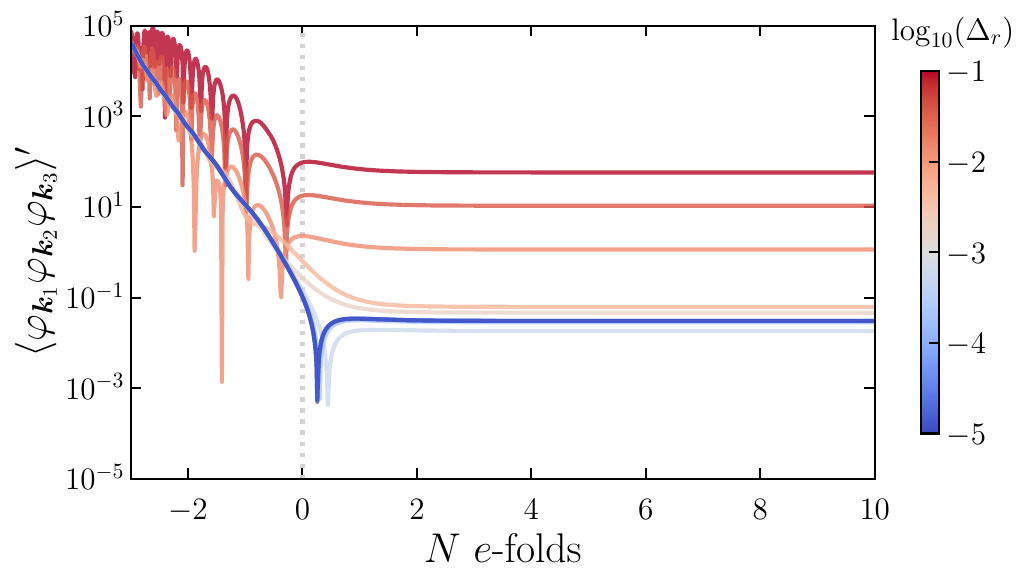}
  \vspace*{0.2cm}
  \caption{Cosmological flow of the three-point correlation function in the equilateral configuration $k_1=k_2=k_3 \equiv k$ as function of the relative tolerance $\Delta_r$ of the numerical integrator (in log scale). Not shown in this figure is the initialisation with the numerical $i\epsilon$ prescription. We have taken $\Delta N=4$ $e$-folds before horizon crossing at $N=0$. [\href{https://github.com/deniswerth/CosmoFlow/blob/main/CosmoFlow/Massless_dphi3/PossibleIssues.ipynb}{\faGithub}]}
  \label{fig: Masslessdphi3 relative error}
\end{figure}

\paragraph{Absolute and relative error.} Numerically solving the flow equations for a sufficiently complicated theory can be challenging and requires good accuracy. The absolute tolerance $\Delta_a$ should be set as low as possible as correlators involving massive fields or conjugate momenta decay outside the horizon until reaching extremely low values. Changing absolute tolerance does not affect the computational speed for the flow equation resolution. On the other hand, there is no general rule to choose the relative tolerance $\Delta_r$, and the most important criterion to chose it is convergence of the final result, i.e.~by decreasing $\Delta_r$ (hence increasing numerical precision) you should see the correlators (typically the one including only massless fields as they usually freeze on super-horizon scales) converge to some value. The integrator has converged if the solution does not vary when $\Delta_r$ is slightly modified. As an illustration, in Fig.~\ref{fig: Masslessdphi3 relative error}, we depict the cosmological flow of the three-point correlation function of fields in the massless $\dot{\varphi}^3$ theory, from deep inside the horizon to super-horizon scales when the modes freeze, varying the relative tolerance. Notice that the solution starts to converge around $\Delta_r \approx 10^{-4}$, and has completely converged for $\Delta_r = 10^{-5}$. More complicated theories require lower relative tolerance to be able to resolve complex physical phenomena, typically around horizon crossing. However, the lower $\Delta_r$ is taken, the slower the code will run. In the end, choosing the correct relative tolerance, finding balance between accuracy and computational speed is a well-recognised challenge for those familiar with numerically solving differential equations.

\vskip 4pt
As the numerical integrator precision is lowered, the poor quality of the solution is readily visible in the appearance of oscillations in the sub-horizon regime. These oscillations are common (yet signals a wrong numerical implementation and/or precision) and typically arise when there is a lack of precision for \textit{cancelling positive and negative frequency modes} in correlators. In the insert below, we illustrate this feature with a simplified example. 

\begin{framed}
{\small \noindent {\it Sub-horizon oscillations.}---Some numerical parameters lead to quickly oscillating correlators inside the horizon. These oscillations of course should be absent in the exact result and are a consequence of a poor numerical precision when cancelling positive and negative frequency modes, as it should be to obtain smooth correlators. Let us consider the following integral
\begin{equation}
    \I_k(\tau) = g \int_{-\infty(1-i\epsilon)}^\tau \d\tau' \tau'^n e^{ik(\tau' - \tau)}\,,
\end{equation}
with $n$ a positive integer, which serves as an illustrating and simplified example of a correlator. We have indicated the $i\epsilon$ prescription in the integral lower bound, meaning that the contour is deformed to (positive) imaginary conformal times in the infinite past to make the integral converge. We are ultimately interested in the correlator evaluated at $\tau=0$. Via an appropriate contour rotation and setting $z = i\tau'$ (with $\tau = 0$), the direct calculation of this integral can be performed on the imaginary axis
\begin{equation}
\label{eq: iepsilon simple example}
    \I_k(0) = -g \int_0^{+\infty}\d (iz) (iz)^n e^{-k z} = -i^{n+1} \,\frac{g}{k^{n+1}} \,\Gamma(n+1)\,.
\end{equation}
Let us now mimic the strategy of the cosmological flow to compute these type of integrals. Differentiating with respect to conformal time $\tau$, we find that the integral $\I_k(\tau)$ should be a solution to the following differential equation
\begin{equation}
    \partial_\tau \I_k(\tau) = -ik \I_k(\tau) + g \tau^n\,.
\end{equation}
The general solution to this equation at all times is
\begin{equation}
\label{eq: iepsilon general solution}
    \I_k(\tau) = \mathcal{A} e^{-ik\tau} - i^{n+1}\, \frac{g}{k^{n+1}}\, e^{-ik\tau} \Gamma(n+1, ik\tau)\,,
\end{equation}
where $\mathcal{A}$ is a constant of integration, and $\Gamma(n, x)$ is the incomplete gamma function. Deep inside the horizon $|k\tau|\gg1$, the first exponential $e^{-ik\tau}$ oscillates with a high frequency and the $i\epsilon$ prescription turns it off adiabatically. This forces $\mathcal{A} = 0$ for consistency. This means that if the correlators are not properly and exactly initialised on the correct attractor solution selecting e.g.~the Bunch Davies initial state, they will pick up this oscillating exponential and oscillate under the horizon. It is therefore crucial to have the numerical $i\epsilon$ prescription activated. 

\vskip 4pt
The entire time dependence of the correlator is therefore encoded in the particular solution. At late times $|k\tau|\rightarrow 0$, we recover the exact result in Eq.~(\ref{eq: iepsilon simple example}). However, at early times $|k\tau|\gg1$, we have 
\begin{equation}
    \Gamma(n+1, ik\tau) \approx e^{ik\tau} (-ik\tau)^n\,.
\end{equation}
We therefore need a precise cancellation between this positive frequency mode and the negative frequency mode present in (\ref{eq: iepsilon general solution}). If the numerical precision is not good enough, this will generate oscillating signals that would contaminate the final result.

}
\end{framed}

\begin{figure}[h!]
  \centering
  \includegraphics[width=0.6\textwidth]{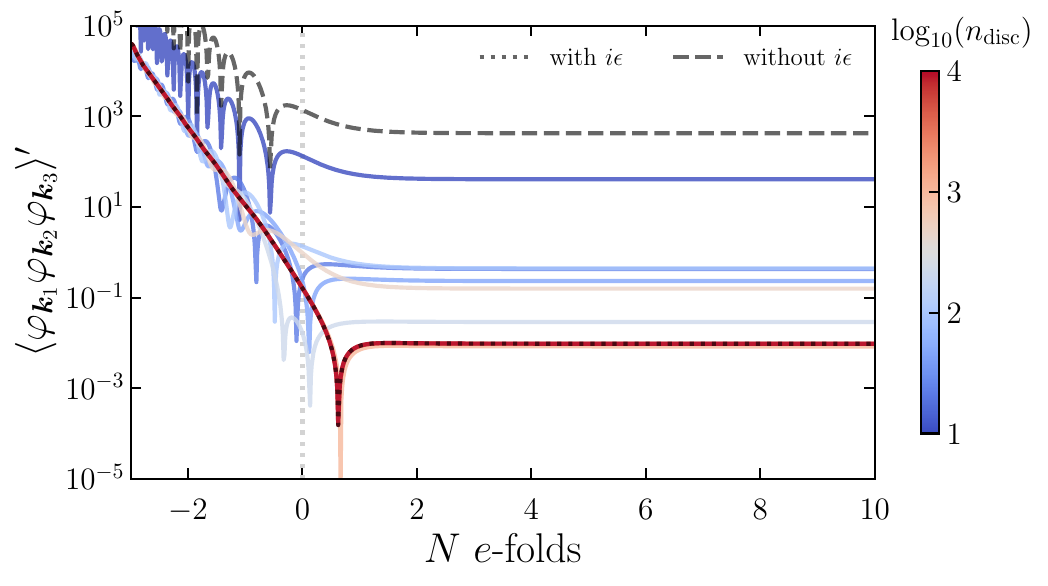}
  \vspace*{0.2cm}
  \caption{Cosmological flow of the three-point correlation function in the equilateral configuration $k_1=k_2=k_3 \equiv k$ as function of the number of discrete points $n_{\text{disc}}$ defining the time grid for continuous functions (in log scale). We have taken $\Delta N=4$ massless $e$-folds before horizon crossing at $N=0$ and have set $\Delta_r = 10^{-3}$. The dotted and dashed lines correspond to the flow when initialising the correlators with and without the numerical $i\epsilon$ prescription, respectively, and for $n_{\text{disc}} = 10^4$ in both cases. [\href{https://github.com/deniswerth/CosmoFlow/blob/main/CosmoFlow/Massless_dphi3/PossibleIssues.ipynb}{\faGithub}]}
  \label{fig: Masslessdphi3 time discretisation}
\end{figure}

\paragraph{Time discretisation and numerical $i\epsilon$ prescription.} It is important to have enough points in the discretisation of time (here in $e$-folds) to properly integrate the flow equations. This is illustrated in Fig.~\ref{fig: Masslessdphi3 time discretisation} where we show the cosmological flow of the three-point correlator varying the number of discretised points for the numerical integration $n_{\text{disc}}$. Below $n_{\text{disc}} \approx 10^3$, the solution develops sub-horizon oscillations and the numerical result is not reliable. These oscillations are exactly the same as the ones in Fig.~\ref{fig: Masslessdphi3 relative error} and their origin is discussed in the insert above. When lowering $n_{\text{disc}}$, the numerical solution slightly gets off the attractor solution at a certain time. The Bunch-Davies state is no longer picked and excited states are visible in the form of negative frequency modes. Cancelling these contaminating oscillations is exactly the purpose of the numerical $i\epsilon$ prescription, as we illustrate in Fig.~\ref{fig: Masslessdphi3 relative error} with the dotted and dashed lines. Without it, the initial flow is deviated from the correct vacuum state and the solution picks up oscillations. On the contrary, correctly implementing the $i\epsilon$ prescription selects the correct vacuum state and the correlators appear smooth. For both curves, we have set $n_{\text{disc}} = 10^4$.

\begin{figure}[h!]
  \centering
  \includegraphics[width=0.6\textwidth]{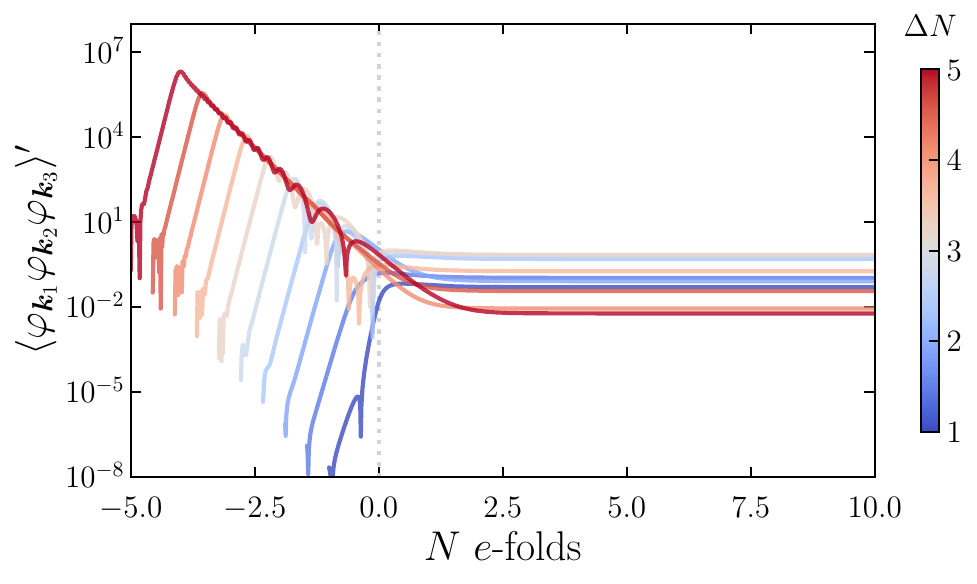}
  \vspace*{0.2cm}
  \caption{Cosmological flow of the three-point correlation function in the equilateral configuration $k_1=k_2=k_3 \equiv k$ as function of the number of massless $e$-folds before horizon crossing at $N=0$. We have taken $n_{\text{disc}} = 10^4$ and $\Delta_r = 10^{-3}$. [\href{https://github.com/deniswerth/CosmoFlow/blob/main/CosmoFlow/Massless_dphi3/PossibleIssues.ipynb}{\faGithub}]}
  \label{fig: Masslessdphi3 massless efolds}
\end{figure}

\paragraph{Massless $e$-folds.} In order to select the correct initial vacuum state for the correlators, they must be initialised sufficiently deep inside the horizon. Typically, around $\Delta N \approx 4$ is enough for the solution to be stable in simple theories. However, note that too many sub-horizon $e$-folds makes the code slow as integrating the flow equations under the horizon is computationally expensive (one must resolve the tiny oscillations resulting from a small error in the cancellation of positive and negative frequency modes). In addition, as illustrated in Fig.~\ref{fig: Masslessdphi3 massless efolds}, too many sub-horizon $e$-folds can destabilise the numerical solution around horizon crossing. Finding an accurate $\Delta N$ for specific runs is done empirically. It usually depends on the theory (the more complex the theory, the more sub-horizon $e$-folds are needed to obtain accurate results), and on the dynamics of both massless fields (e.g.~in the presence of a reduced speed of sound $c_s\ll 1$, modes freeze at $|k\tau|=c_s^{-1}$ before the usual horizon crossing, and the number of massless $e$-folds should be increased) and massive fields which start to decay around mass-shell crossing $|k\tau| \sim m/H$. Our advice is to perform tests to choose the correct numerical parameters before launching longer and complicated computations. Lastly, in theories with non-trivial super-horizon evolution, one must ensure the convergence of the correlators that are conserved, typically of massless fields or those of curvature perturbation $\zeta$.

\section{Applications: Cosmological Correlators in all their Forms}
\label{sec: Applications}

Let us now guide the reader through various concrete applications that showcase what \textsf{CosmoFlow} can compute. The following worked examples serve both to illustrate the numerous possibilities that this code provides, and to show the user how to implement different calculations. Therefore, all the following case studies and figures can be reproduced with ready-to-use \href{https://jupyter.org/}{\textsf{Jupyter}} notebooks that can be found on the \textsf{CosmoFlow} \href{https://github.com/deniswerth/CosmoFlow/tree/main}{GitHub repository}. Throughout the examples, we will provide links to the relevant notebooks using the GitHub icon \faGithub.

\subsection{Time Evolution}

Relying on solving differential equations in time, the first evident (yet extremely useful) application of \textsf{CosmoFlow} is that it allows for a direct access to the time evolution of cosmological correlators, from their emergence in the infinite past (typically well inside the horizon) to later times (typically well outside the horizon). The notebook to generate figures of this section can be found here: \href{https://github.com/deniswerth/CosmoFlow/blob/main/CosmoFlow/PhiPsi/TimeEvolution.ipynb}{\faGithub}.

\paragraph{Two-field theory.} As a more complicated example, we consider the theory of a massless scalar field $\varphi$ interacting with a massive scalar field $\Psi$ in de Sitter. These fields can interact quadratically, allowing both particles to mix at the linear level, and at the cubic level so that the theory can generate cubic correlations among the fields. The theory we consider is the following:

\begin{equation}
\label{eq: phi-psi Lagrangian}
    \begin{aligned}
    S = \int \d t\d^3x\, a^3&\left[\,\frac{1}{2}\dot{\varphi}^2 - \frac{c_s^2}{2}\frac{(\partial_i \varphi)^2}{a^2} - \frac{1}{2}\left[(\partial_\mu \Psi)^2 + m^2\Psi^2\right] + \rho\, \dot{\varphi}\Psi \right.\\
    &\left.- \frac{\lambda_1}{2} \frac{(\partial_i \varphi)^2}{a^2}\,\Psi - \frac{\lambda_2}{2}\, \dot{\varphi} \,\Psi^2 - \frac{\lambda_3}{3!}\, \Psi^3 \right]\,,
    \end{aligned}
\end{equation}
where overdots denote derivatives with respect to cosmic (physical) time, $c_s$ is the speed of sound of $\varphi$---allowing a breaking of de Sitter boosts---, $m$ is the mass of $\Psi$, $\rho$ is the linear mixing strength, and $\lambda_i$ with $i=1, 2, 3$ are arbitrary coupling constants. The first line corresponds to the quadratic theory and the second line contains cubic interactions. Such theory typically arises when the Goldstone boson of broken time translations during inflation is coupled to a scalar massive field, see~\cite{Pinol:2023oux} for a detailed study of such theories. The quadratic theory leads to linear equations of motions whose analytical solutions are unknown. However, at weak mixing (typically when $\rho \ll H$ where $H$ is the Hubble scale), the linear mixing can be treated perturbatively and (some) analytical computations are tractable. In the limit $\rho/H \rightarrow 0$, both sectors decouple at the linear level and the flow equations become block diagonal. 

\vskip 4pt
After deriving the linear conjugate momenta and performing a Legendre transform, the identifications of the different tensors for the numerical implementation leads to
\begin{equation}
\begin{aligned}
    \Delta_{\alpha\beta} &= \delta_{\alpha\beta}\,, \quad M_{\alpha\beta} = 
    \begin{pmatrix}
    -c_s^2 \tfrac{k^2}{a^2} & 0 \\
    0 & -\left(\tfrac{k^2}{a^2} + m^2 + \rho^2\right)
    \end{pmatrix}
    \,, 
    \quad I_{\alpha\beta} = 
    \begin{pmatrix}
    0 & \rho \\
    0 & 0
    \end{pmatrix}\,,\\ A_{\varphi\varphi\Psi} &= \lambda_1 \frac{\bm{k}_\varphi \cdot \bm{k}_\varphi}{a^2}\,, \quad A_{\Psi\Psi\Psi} = -\frac{\lambda_3}{3} + \rho\lambda_2\,, \quad B_{\Psi\Psi\varphi} = -\lambda_2\,.
\end{aligned}
\end{equation}
When implementing these tensors in the \textsf{Theory.py} module, one should be careful to properly symmetrise the tensors. Moreover, we recall that the Fourier summation notation in~(\ref{eq: Hamiltonian}) contains a momentum conserving delta function that flips the sign of a momentum in the case of spatial gradient interactions. Note that slight modifications have been made in the implementation of the initial conditions to take into account the possibility of a reduced speed of sound, see the script \href{https://github.com/deniswerth/CosmoFlow/blob/main/CosmoFlow/PhiPsi/Solver.py}{\faGithub} for more details.

\begin{figure}[h!]
  \centering
  \includegraphics[width=0.6\textwidth]{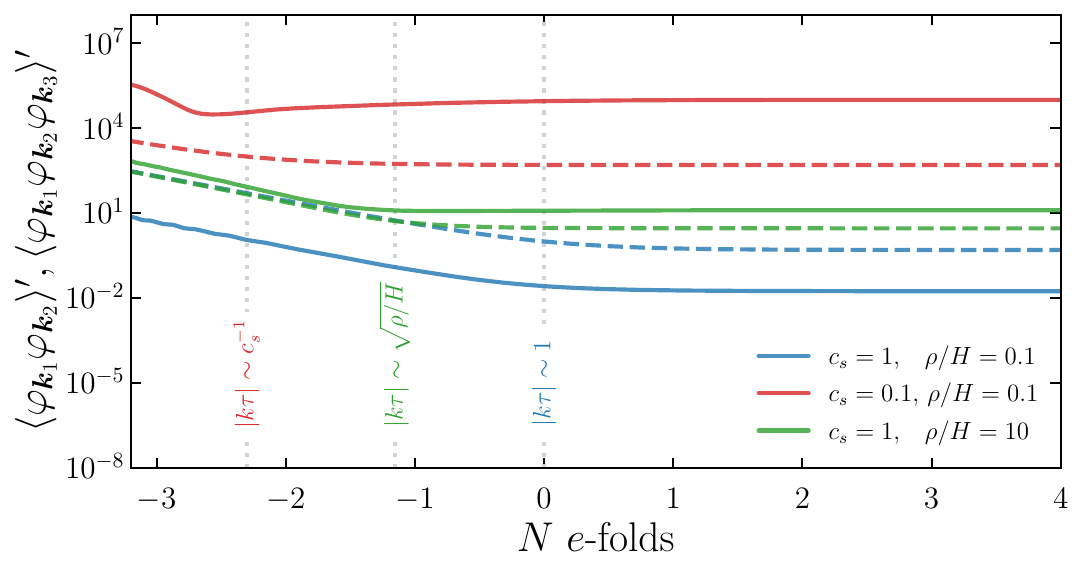}
  \vspace*{0.2cm}
  \caption{Cosmological flow of the two-point correlators $\braket{\varphi_{\bm{k}_1} \varphi_{\bm{k}_2}}'$ (dashed lines) and the three-point correlators $\braket{\varphi_{\bm{k}_1} \varphi_{\bm{k}_2} \varphi_{\bm{k}_3}}'$ (solid lines), for the interaction $(\partial_i \varphi)^2\Psi$ whose size is governed by $\lambda_1$, in the equilateral kinematic configuration $k_1=k_2=k_3$ for $c_s=1, \rho/H=0.1$ (in \textcolor{pyblue}{blue}), $c_s=0.1, \rho/H=0.1$ (in \textcolor{pyred}{red}), and $c_s=1, \rho/H=10$ (in \textcolor{pygreen}{green}). The mass of the field $\Psi$ is taken to be $m/H=2$. Not shown in this figure are all other correlators, mixing the fields $\varphi$ and $\Psi$, as well as their conjugate momenta. We have taken $\Delta N=4.5$ massless $e$-folds before horizon crossing at $N=0$, $n_{\text{disc}} = 10^4$ and $\Delta_r = 10^{-4}$ as numerical parameters. [\href{https://github.com/deniswerth/CosmoFlow/blob/main/CosmoFlow/PhiPsi/TimeEvolution.ipynb}{\faGithub}]}
  \label{fig: PhiPsi different horizons}
\end{figure}

\paragraph{Characteristic times.} The quadratic theory cannot be solved analytically in the entire parameter space, i.e.~for all ranges for the speed of sound $c_s$, the mass $m$ of the additional field $\Psi$, and the strength of the linear mixing between both fields $\rho$, and analytical solutions for three-point correlators are out of reach. Therefore, this example showcases how one can use \textsf{CosmoFlow} to gain intuition when studying complex theories. In Fig.~\ref{fig: PhiPsi different horizons}, we show the cosmological flow of the two- and three-point correlators of the field $\varphi$ for three different cases. For a unity speed of sound $c_s=1$ and a weak mixing $\rho/H \ll 1$ (in Hubble units), correlators of the massless field $\varphi$ freeze at the usual \textcolor{pyblue}{\textit{Hubble horizon}} given by $|k\tau| \sim 1$, here at $N=0$. At weak mixing, when the speed of sound is reduced, the massless mode freezes at the \textcolor{pyred}{\textit{sound horizon}} $|k\tau|\sim c_s^{-1}$, before the usual Hubble horizon, and with a globally greater amplitude. This is readily visible in Fig.~\ref{fig: PhiPsi different horizons} where various horizons are denoted with vertical dotted lines. At strong mixing, the coupled dynamics is drastically modified and massless modes freeze at the \textcolor{pygreen}{$\rho$\textit{-horizon}} $|k\tau|\sim \sqrt{\rho/H}$, here for $c_s=1$. Note that for $c_s=1$, the weak and strong mixing cases lead to the same two-point correlators deep inside the horizon. This is expected as the system asymptotically reaches the Bunch-Davies state in the infinite past, and the two fields are effectively decoupled. However, the mixing between both fields builds up progressively around $\rho$ and Hubble crossing, leading to an amplification of the correlators in the strong mixing case. Similarly, at weak mixing, one can show that the massive field $\Psi$ starts to decay at the \textit{mass-shell horizon} $|k\tau|\sim m/H$. Because \textsf{CosmoFlow} allows for a direct access to the bulk time evolution of cosmological correlators, it can be used to identify characteristic time scales of the theory and guide analytical calculations in limiting regimes, as these times play an important role when computing bulk time integrals.

\begin{figure}[h!]
  \centering
  \includegraphics[width=0.6\textwidth]{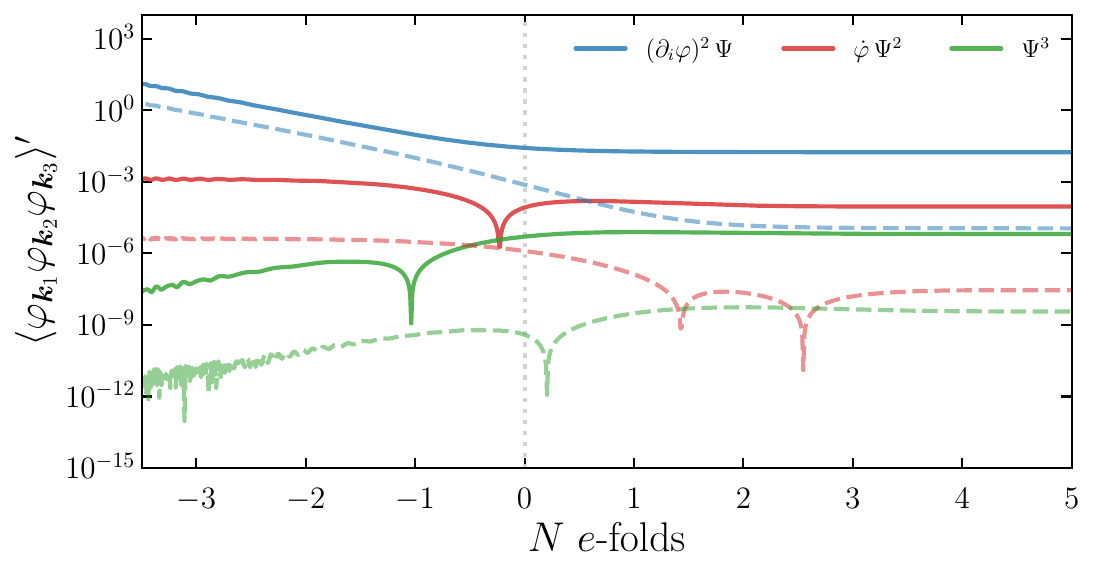}
  \vspace*{0.2cm}
  \caption{Cosmological flow of the three-point correlators $\braket{\varphi_{\bm{k}_1} \varphi_{\bm{k}_2} \varphi_{\bm{k}_3}}'$ in the equilateral configuration $k_1=k_2=k_3\equiv k_\star$ (solid lines) and in a mildly squeezed configuration $k_1=k_2\equiv 5k_\star, k_3=k_\star$ (dashed lines), for the three considered cubic interactions, where $k_\star=1$ is a pivot scale so that the corresponding mode exists the Hubble horizon at $N=0$. We have set $c_s=1, m/H=2, \rho/H=0.1$, and $\lambda_i = 1$ with $i=1, 2, 3$. The numerical parameters are $\Delta N = 5, n_{\text{disc}} = 10^4$ and $\Delta_r = 10^{-5}$. The vertical dotted line denotes Hubble horizon crossing for the mode $k_\star$. [\href{https://github.com/deniswerth/CosmoFlow/blob/main/CosmoFlow/PhiPsi/TimeEvolution.ipynb}{\faGithub}]}
  \label{fig: PhiPsi cubic interactions}
\end{figure}

\paragraph{Cubic interactions.} Computing cosmological correlators is very challenging. At weak mixing $\rho/H\ll 1$, the dominant contributions to the correlator $\braket{\varphi_{\bm{k}_1} \varphi_{\bm{k}_2} \varphi_{\bm{k}_3}}'$ are single-, double-, and triple-diagrams for the interactions $(\partial_i \varphi)^2\Psi$, $\dot{\varphi}\Psi^2$ and $\Psi^3$, respectively. At strong mixing, one needs to resum the infinite chain of linear mixing insertions to obtain the full correlators. Without mixing $\rho/H = 0$, there is zero correlation among three fields $\varphi$. In Fig.~\ref{fig: PhiPsi cubic interactions}, we show the cosmological flow of the tree-point correlator $\braket{\varphi_{\bm{k}_1} \varphi_{\bm{k}_2} \varphi_{\bm{k}_3}}'$ generated by the three cubic interactions considered in Eq.~(\ref{eq: phi-psi Lagrangian}), in two different kinematic configurations. Of course, these interactions lead to different final values for the correlator once the massless modes have frozen. However, less evident are the different scalings in the sub-horizon regime. These behaviours can be understood in the following way. Correlators sourced by interactions containing spatial gradients are amplified by the scale factor under the horizon, as the physical momentum reads $k_{\text{p}} = k/a$. These interactions usually lead to very stable results as the solution quickly reaches the flow equation attractor without too many polluting sub-horizon oscillations. On the other hand, other interactions---such as $\Psi^3$ or including time derivatives (which are associated with conjugate momenta)---are suppressed compared to spatial gradients. These features are well visible in Fig.~\ref{fig: PhiPsi cubic interactions}. These interactions typically lead to less stable numerical results and more precision is required to properly resolve them. 

\vskip 4pt
Varying the kinematic configuration, one sees that correlators in mildly squeezed kinematic configurations freeze later than correlators in the equilateral configuration. This is simply due to the fact that shorter modes exit their Hubble horizon later than long modes, as illustrated in Fig.~\ref{fig: PhiPsi cubic interactions}. After familiarising with the notebook, the user might notice that solving the flow equations for softer kinematic configurations is numerically more challenging and time consuming. This is the result of solving the dynamics of a system with a large hierarchy of scales, and more massless $e$-folds before horizon crossing are resolved. Properly initialising and evolving correlators in the sub-horizon regime requires a lot of precision and is essential to obtain accurate results in complicated theories, e.g~by increasing $n_{\text{disc}}$ and $\Delta N$, and reducing $\Delta_r$. Note, however, that the visible numerical noise in the sub-horizon regime for the correlator generated by the interaction $\Psi^3$ in the mildly squeezed configuration is harmless as the correlator becomes smooth well before crossing the horizon.

\begin{framed}
{\small \noindent {\it Exercise.}---We invite the reader to familiarise with \textsf{CosmoFlow} through an exercise that can be completed in roughly 2h, depending on the new user's relentlessness. Let us study the dynamics of the low-energy mode associated with the spontaneous breaking of a continuous global symmetry in an expanding background. The theory we consider contains a massless Goldstone boson and a massive excitation that mix at the linear order through a chemical potential. Here, using \textsf{CosmoFlow}, you are invited to study the accuracy of a simplified theory in which the heavy mode has been integrated out. 

\vskip 4pt
We consider a relativistic theory of a complex scalar field $\Phi$ in de Sitter that has a global $U(1)$ symmetry
\begin{equation}
    \mathcal{L}/a^3 = -(\partial_\mu \Phi^\dagger)(\partial^\mu \Phi) - M^2\, \Phi^\dagger\Phi - \lambda\, (\Phi^\dagger\Phi)^2\,,
\end{equation}
where $a(t) = e^{Ht}$ with $H$ being the constant Hubble scale, and $M, \lambda$ constant parameters. This Lagrangian is a textbook example of a renormalisable (i.e.~UV-complete) theory with a global symmetry (usually studied in flatspace). 

\begin{itemize}
    \item\textbf{Theory.} Write the complex field as $\Phi = \tfrac{r}{\sqrt{2}}e^{i\theta/v}$ in terms of an angular field $\theta(t, \bm{x})$ and a radial part $r(t, \bm{x})$, where $v$ is some energy scale, and derive the corresponding two-field Lagrangian. We are interested in studying the theory around a time-dependent background $\theta \propto t$. Introduce the Goldstone boson $\pi(t, \bm{x})$ such that $\theta(t, \bm{x}) = \mu^2 t/2 + \pi(t, \bm{x})$ where $\mu$ is the chemical potential (we can assume it to be positive without loss of generality), and derive the Lagrangian that couples $\pi$ to $r$. Expanding the radial field around the minimum $r(t, \bm{x}) = v + \sigma(t, \bm{x})$, shows that the Lagrangian boils down to
    \begin{equation}
    \label{eq: pi-sigma theory}
        \mathcal{L}/a^3 = -\frac{1}{2}(\partial_\mu \pi)^2 - \frac{1}{2}(\partial_\mu \sigma)^2 - \frac{1}{2}m^2\sigma^2 + \frac{\mu^2}{2v^2}\left(2v\sigma + \sigma^2\right)\dot{\pi} - \frac{1}{v}(\partial_\mu\pi)^2\sigma - \lambda v \sigma^3\,,
    \end{equation}
    with $m^2 = 2\lambda v^2$, and where we have discarded quartic interactions and constant terms.
    
    \item \textbf{Numerical implementation.} After switching to the corresponding Hamiltonian, introducing the conjugate momenta $p_\pi$ and $p_\sigma$ to the fields, derive the tensors $\Delta_{\upalpha\upbeta}$, $M_{\upalpha\upbeta}$, $I_{\upalpha\upbeta}$, $A_{\upalpha\upbeta\upgamma}$, $B_{\upalpha\upbeta\upgamma}$, $C_{\upalpha\upbeta\upgamma}$ and $D_{\upalpha\upbeta\upgamma}$, and implement the theory in \textsf{CosmoFlow}. For this, you can for example create a new folder in your \textsf{CosmoFlow} repository.
    \item \textbf{Effective theory.} When the massive field $\sigma$ is heavy enough, it can be integrated out. At tree-level, derive the effective single-field Lagrangian for $\pi$ by setting the massive field on-shell at leading-order in gradient expansion, i.e.~$\sigma \approx \tfrac{\mu^2}{m^2 v}\dot{\pi}$. Implement this new single-field theory in \textsf{CosmoFlow}. Compute the three-point correlator $\braket{\pi_{\bm{k}_1} \pi_{\bm{k}_2} \pi_{\bm{k}_3}}'$ in the equilateral kinematic configuration $k_1=k_2=k_3$ as a function of the mass $m/H\in[10^{-1}, 10]$ (use a log-scale discretisation) in both theories. Fixing $v/H=1$ and $m/H=3$, generate a figure that shows the rescaled three-point correlator $(k_1k_2k_3)^2\braket{\pi_{\bm{k}_1} \pi_{\bm{k}_2} \pi_{\bm{k}_3}}'$ in the isosceles kinematic configuration $k_1=k_2\equiv k_{\text{S}}$ and $k_3 \equiv k_{\text{L}}$ as a function of the squeezing ratio $k_{\text{L}}/k_{\text{S}} \in [10^{-2}, 1]$, for both theories. What are your conclusions?
\end{itemize}
}
\end{framed}

\subsection{Theory Dependence} 
\label{subsec: Theory Dependence}

\textsf{CosmoFlow} offers the possibility to easily navigate through a vast landscape of theories, e.g.~varying coupling constant strengths. Indeed, contrary to C++, Python does not require compilation before execution. Moreover, the module \textsf{Theory.py} allows to pre-implement a large class of theories at the same time. This way, the structure of \textsf{CosmoFlow} makes it possible to scan different theories without the need to ever modify the source code. In this section, we illustrate how one can use the code to scan over theories with two examples: (i) scanning over parameters of the theory, and (ii) implementing various time dependencies for couplings.

\begin{figure}[h!]
  \centering
  \subfloat{\includegraphics[width=0.45\textwidth]{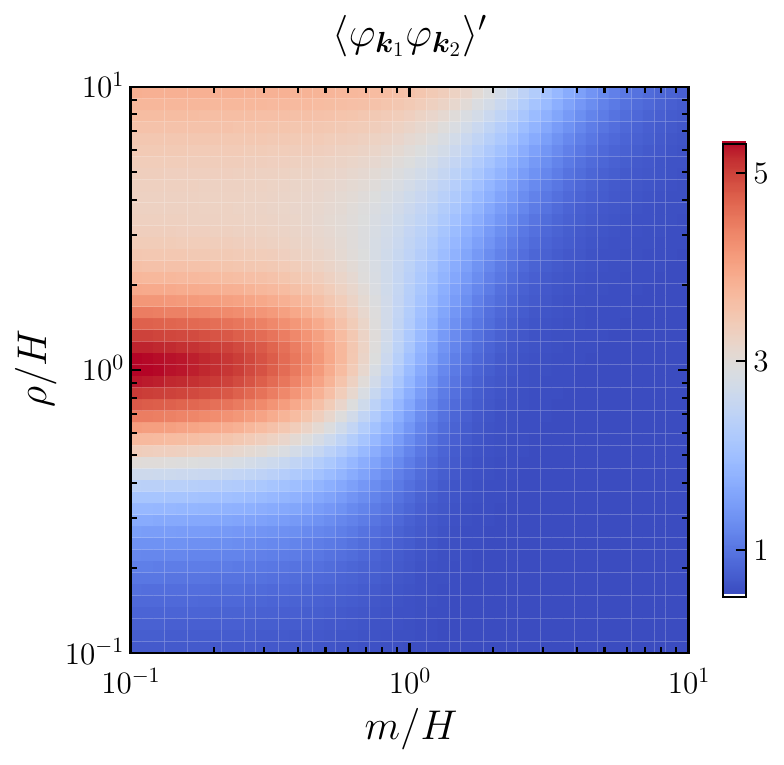}}
  \hspace*{0.2cm}
  \subfloat{\includegraphics[width=0.45\textwidth]{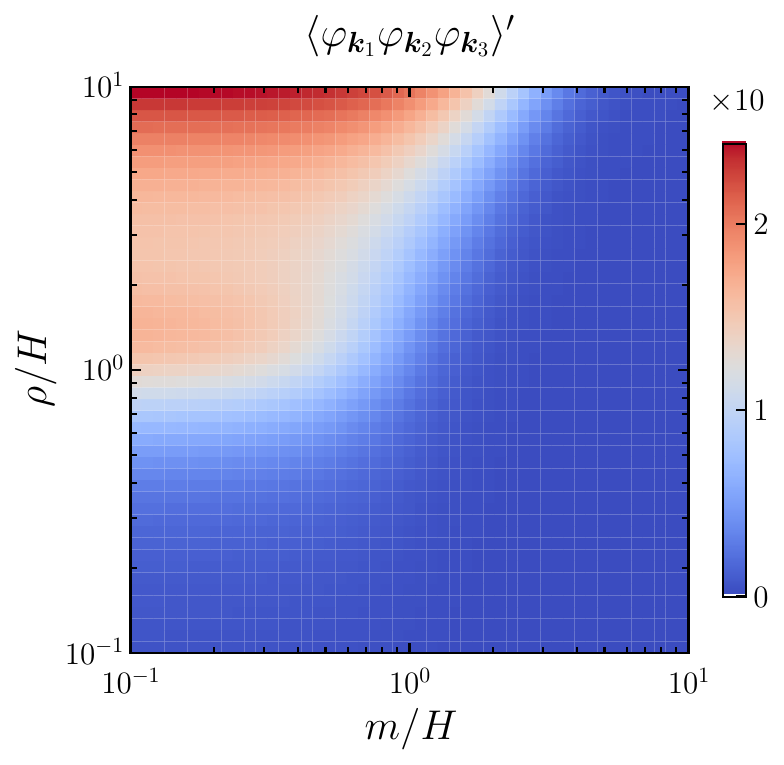}}
  \vspace*{0.2cm}
  \caption{\textit{Left}: Phase diagram of two-point correlation function $\braket{\varphi_{\bm{k}_1} \varphi_{\bm{k}_2}}'$ as function of the mass $m/H$ of the field $\Psi$ and the linear mixing strength $\rho/H$ (in Hubble units), see~(\ref{eq: phi-psi Lagrangian}). \textit{Right}: Phase diagram of the three-point correlation function $\braket{\varphi_{\bm{k}_1} \varphi_{\bm{k}_2} \varphi_{\bm{k}_3}}'$ in the equilateral configuration $k_1=k_2=k_3$ for the interaction $(\partial_i \varphi)^2 \Psi$. For both diagrams, we have set $\Delta_r = 10^{-4}, n_{\text{disc}} = 10^4$ and $\Delta N = 4.5$ of massless $e$-folds. The asymptotic value as $\tau\rightarrow 0$ is taken to be $5$ $e$-folds after horizon crossing. [\href{https://github.com/deniswerth/CosmoFlow/blob/main/CosmoFlow/PhiPsi/PhaseDiagram.ipynb}{\faGithub}]}
  \label{fig: Phase Diagram}
\end{figure}

\paragraph{Phase diagram.} Fixing $c_s=1$, the quadratic theory given in Eq.~(\ref{eq: phi-psi Lagrangian}) only depends on two parameters, the mass $m$ of the field $\Psi$ and the linear mixing strength $\rho$. Therefore, these two parameters define a two-dimensional phase diagram for the correlators. In Fig.~\ref{fig: Phase Diagram}, we show the two- and three-point correlators of $\varphi$ in the $(m/H, \rho/H)$ plane, $5$ $e$-folds after Hubble crossing. Various regimes of this phase diagram are discussed in~\cite{Pinol:2023oux}. Note that at weak mixing and light massive field case, the latter continues sourcing the field $\varphi$ after horizon crossing, creating non-trivial super-horizon evolution for the correlators. As such, the choice to take $5$ $e$-folds of super-horizon evolution is not enough to account for the complete dynamics in the entire phase diagram. The corresponding notebook to generate this figure can be found on the following link: \href{https://github.com/deniswerth/CosmoFlow/blob/main/CosmoFlow/PhiPsi/PhaseDiagram.ipynb}{\faGithub}. Scanning on parameters of the theory requires solving multiple times the flow equations, i.e.~for every set of parameters, and can be time consuming. However, the cosmological flow method has several advantages. First, the structure of the flow equations forming a closed system allows for all two- and three-point correlation functions to be solved at once. Therefore, for a fixed set of theory parameters, a single \textsf{CosmoFlow} run allows to compute all correlators. Second, each resolution of the flow equations operates independently, which allows for parallelisation. In the ready-to-use notebook, we showcase this procedure with the Python package \href{https://joblib.readthedocs.io/en/stable/}{\textsf{joblib}}, that can be installed with \textsf{pip}. Naturally, any other parallelisation methods can be used. Our advice is to follow the correct parallelisation implementation using the \textsf{top} command on the terminal, e.g.~checking that several cores are used for the computation. In general, parallelisation allows to gain roughly a factor $10$ on the runtime, and is regularly used.

\begin{figure}[h!]
  \centering
  \subfloat{\includegraphics[width=0.45\textwidth]{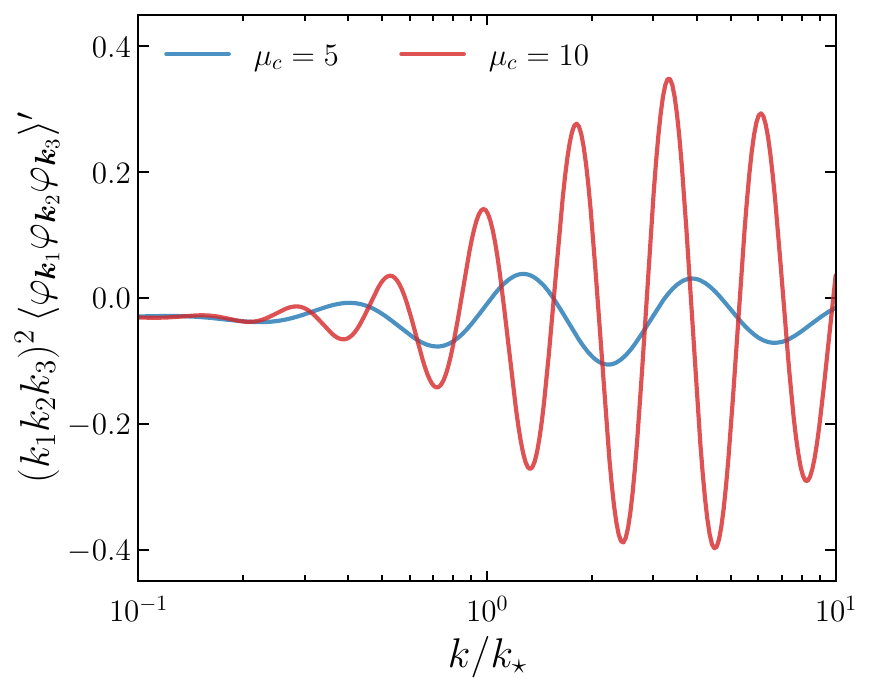}}
  \hspace*{0.2cm}
  \subfloat{\includegraphics[width=0.45\textwidth]{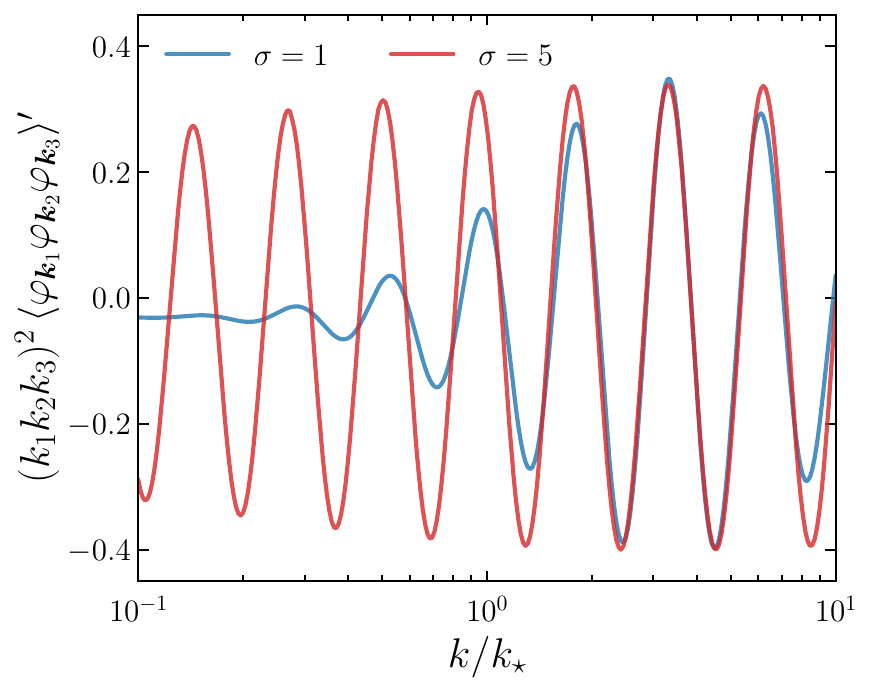}}
  \vspace*{0.2cm}
  \caption{\textit{Left}: Rescaled three-point correlation function $\braket{\varphi_{\bm{k}_1} \varphi_{\bm{k}_2} \varphi_{\bm{k}_3}}'$ in the equilateral configuration $k_1=k_2=k_3=k$ as a function of $k/k_\star$ where $k_\star$ is the scale at which the feature in $g(t)$ is localised, i.e.~corresponding to the time $N_\star = H t_\star = \log(k_\star)$, for $\mu_c = 5$ and $\mu_c = 10$. We have chosen $g_0 = 1, \delta g = 0.1, Ht_\star \equiv N_\star = 0$ and $\sigma = 1$. \textit{Right}: Same rescaled three-point correlation function $\braket{\varphi_{\bm{k}_1} \varphi_{\bm{k}_2} \varphi_{\bm{k}_3}}'$ as a function of $k/k_\star$, for $\sigma = 1$ and $\sigma = 5$. We have chosen $g_0 = 1, \delta g = 0.1, Ht_\star \equiv N_\star = 0$ and $\mu_c = 10$. The numerical parameters are $\Delta_r = 10^{-4}, n_{\text{disc}} = 10^5$ and $\Delta N = 4$. [\href{https://github.com/deniswerth/CosmoFlow/blob/main/CosmoFlow/Massless_dphi3/TimeDependentCouplings.ipynb}{\faGithub}]}
  \label{fig: time dependent couplings}
\end{figure}

\paragraph{Time-dependent couplings.} Let us now illustrate how \textsf{CosmoFlow} deals with time varying coupling constants. Such cases typically arise in concrete inflationary models where the backgrounds fields---whose time evolution feeds the coupling constant time dependence at the level of fluctuations---have non-trivial trajectories, and require numerical evaluation. Within the $\dot{\varphi}^3$ theory in de Sitter~(\ref{eq: dphi3 action}), we consider that the cubic coupling $g(t)$ acquires the following time dependence
\begin{equation}
    g(t) = g_0 + \delta g \times \cos\left(\mu_c t\right) \times \text{exp}\left[-(t - t_\star)^2/2(\sigma/H)^2\right]\,,
\end{equation}
where $\mu_c = \omega_c/H$ is the oscillatory frequency in Hubble units. Such time dependence describes a small oscillatory signal with amplitude $\delta g$ localised around $t_\star$ and with width $\sigma$, on top of a constant coupling strength with amplitude $g_0$. In Fig.~\ref{fig: time dependent couplings}, we show the rescaled three-point correlator in the equilateral kinematic configuration as a function of the scale $k/k_\star$ where $k_\star \equiv e^{Ht_\star}$, varying the frequency and the width of the feature. This quantity would be closely related to the usual bispectrum shape function whose amplitude is usually denoted $f_{\text{NL}}$. The figure can be reproduced using the following notebook: \href{https://github.com/deniswerth/CosmoFlow/blob/main/CosmoFlow/Massless_dphi3/TimeDependentCouplings.ipynb}{\faGithub}. Because both the initial integration time and the time at which the corresponding mode $k/k_\star$ exists the horizon change, note that the \textsf{adiabatic} function that implements the $i\epsilon$ prescription has been adapted accordingly. Each curve necessitates around $\sim2$min of runtime on a Macbook Pro with M1 CPU and running MacOS 11.6 Big Sur, with $300$ points for the discretisation of $k/k_\star$ to properly resolve the oscillations. We also have parallelised the computation to speed up the numerical resolution.

\subsection{Kinematic Dependence}
\label{subsec: Kinematic Dependence}

To further demonstrate the utility of the cosmological flow numerical implementation, we now show how one can study the kinematic dependence of cosmological correlators. This application is of great importance when studying non-Gaussian phenomenology of inflationary theories, see e.g.~\cite{Maldacena:2002vr, Creminelli:2003iq, Chen:2010xka, Renaux-Petel:2015bja, Meerburg:2019qqi} and the references therein.

\paragraph{Correlator shape.} For concreteness, let us focus on the two-field theory given in Eq.~(\ref{eq: phi-psi Lagrangian}) and study the three-point correlation function $\braket{\varphi_{\bm{k}_1} \varphi_{\bm{k}_2} \varphi_{\bm{k}_3}}'$ generated by the cubic interaction $\dot{\varphi}\Psi^2$, setting $\lambda_2=1$. An analytical closed-form expression for this diagram is unknown. There are several ways to present numerical results. In perfect scale invariant theories, as it is the case for this example, the three-point function does not have any \textit{scale} dependence, in the sense that it is insensitive to the overall size of the triangle formed by the three momenta, e.g.~parametrised by the perimeter $k_1+k_2+k_3$ of the triangle. However, it does depend on the \textit{shape} of the triangle. Without loss of generality, we order the momenta such that $k_1 \leq k_2 \leq k_3$ and fix $k_3$. This way, the shape information only depends on the two dimensionless ratios $k_1/k_3$ and $k_2/k_3$ that satisfy $0 \leq k_1/k_3 \leq k_2/k_3 \leq 1$ and $k_3 \leq k_1 + k_2$ by virtue of the triangle inequality. The range of the dimensionless parameters therefore fall inside a triangle whose vertices correspond to the equilateral configuration ($k_1/k_3 = k_2/k_3 = 1$), the squeezed limit ($k_1/k_3 \ll 1, k_2/k_3 = 1$), and the isosceles folded configuration ($k_1/k_3 = k_2/k_3 = 0.5$). An alternative way to measure the shape is to use the Fergusson \& Shellard parametrisation~\cite{Fergusson:2006pr}. It is convenient to plot the rescaled three-point function $(k_1 k_2 k_3)^2\,\braket{\varphi_{\bm{k}_1} \varphi_{\bm{k}_2} \varphi_{\bm{k}_3}}'$ as it is scale-invariant and is related to the usual bispectrum shape function up to a normalisation~\cite{Babich:2004gb}. In practice, as can be noticed in the notebook, we arbitrarily fix the scale $k_3=1$ by virtue of scale-invariance, and choose the scale factor $a=e^N$ so that this mode exits its horizon at $N=0$. This way, the number of massless $e$-folds $\Delta N$ can be unambiguously defined with respect to $N=0$ (we set $H=1$). To avoid changing $\Delta N$ as we scan over soft limits, these are reached by increasing $k_1\sim k_2 \gg k_3$ (corresponding to the short modes) which therefore exit their horizon later. Of course, by scale invariance, this is completely equivalent to fixing $k_1$ and $k_2$ and decreasing $k_3$. However, the latter method is numerically less practical.

\begin{figure}[h!]
  \centering
  \includegraphics[width=0.5\textwidth]{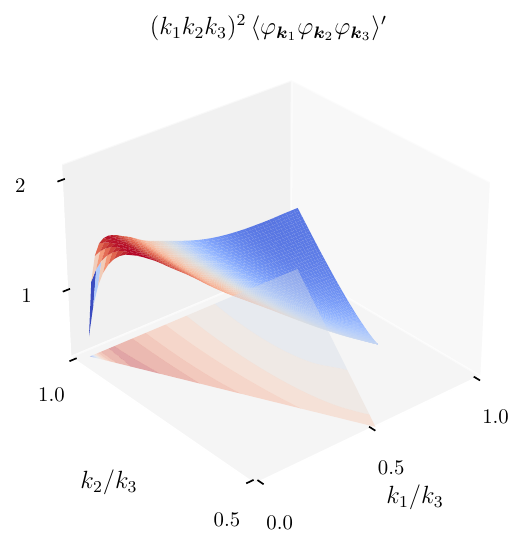}
  \vspace*{0.2cm}
  \caption{The rescaled three-point correlator $\braket{\varphi_{\bm{k}_1} \varphi_{\bm{k}_2} \varphi_{\bm{k}_3}}'$ normalised to unity in the equilateral configuration, for the interaction $\dot{\varphi}\Psi^2$ and $c_s = 0.05, m/H = 2$ and $\rho/H=0.1$. We have set $\Delta N = 4, n_{\text{disc}} = 10^4$ and $\Delta_r = 10^{-3}$. [\href{https://github.com/deniswerth/CosmoFlow/blob/main/CosmoFlow/PhiPsi/KinematicDependence.ipynb}{\faGithub}]}
  \label{fig: shape}
\end{figure}

\vskip 4pt
In Fig.~\ref{fig: shape}, we plot the rescaled three-point correlator in the $(k_1/k_3, k_2/k_3)$ plane, normalised to unity in the equilateral configuration. The corresponding shape is neither faithfully represented by the standard local shape~\cite{Gangui:1993tt, Wang:1999vf, Verde:1999ij, Komatsu:2001rj} nor the equilateral one~\cite{Babich:2004gb, Creminelli:2005hu}. Indeed, the parameters we used, $c_s=0.05, m/H=2$ and $\rho/H=0.1$, fall within the low-speed collider regime, in which the signal is characterised by a resonance in mildly squeezed configurations, see~\cite{Jazayeri:2022kjy, Jazayeri:2023xcj} for more details. Our numerical method is perfectly capable of capturing this effect. Fig.~\ref{fig: shape} can be reproduced with the notebook \href{https://github.com/deniswerth/CosmoFlow/blob/main/CosmoFlow/PhiPsi/KinematicDependence.ipynb}{\faGithub}, and generating a full shape with reasonable accuracy takes around $\sim10$min on a typical laptop. Note that we have parallelised the computation in the notebook as every flow equation integration can be performed independently. Although it might seem that the code is slow from the perspective of pure numerical performance, it is actually very competitive compared to the few full analytical results that are usually expressed in terms of series that converge slowly (or not at all) in particular kinematic configurations, see e.g.~\cite{Arkani-Hamed:2018kmz, Sleight:2019mgd, Sleight:2019hfp, Jazayeri:2022kjy, Pimentel:2022fsc, Qin:2022fbv, Qin:2023ejc, Xianyu:2023ytd}. Certainly, the optimal approach involves designing simple templates capable of instantaneous evaluation. Such templates can be effectively used, for instance, in performing parameter estimations using Markov Chain Monte Carlo methods. In this respect, \textsf{CosmoFlow} can help guiding the search for new templates.

\begin{figure}[h!]
  \centering
  \subfloat{\includegraphics[width=0.45\textwidth]{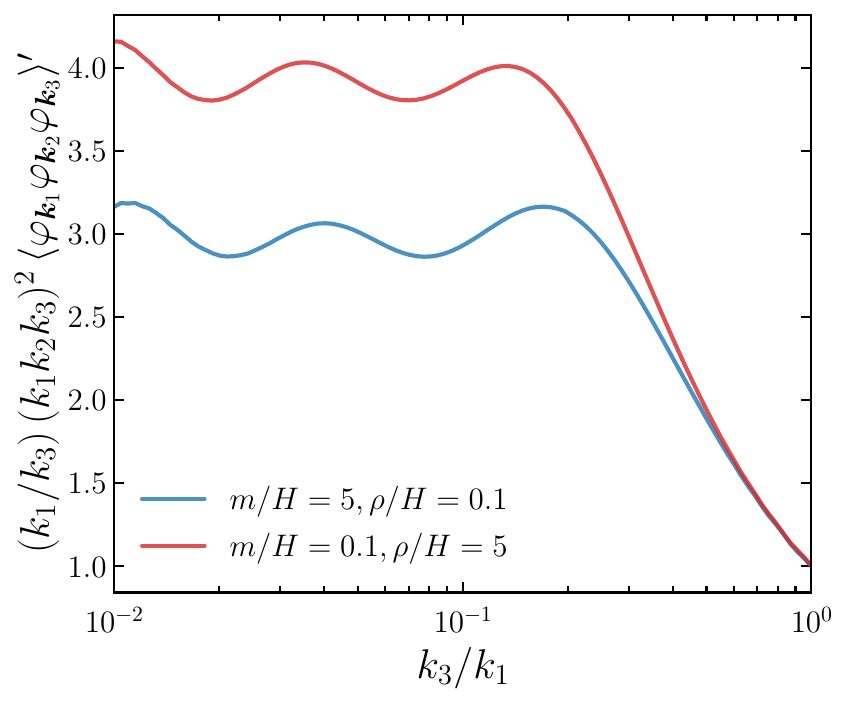}}
  \hspace*{0.2cm}
  \subfloat{\includegraphics[width=0.45\textwidth]{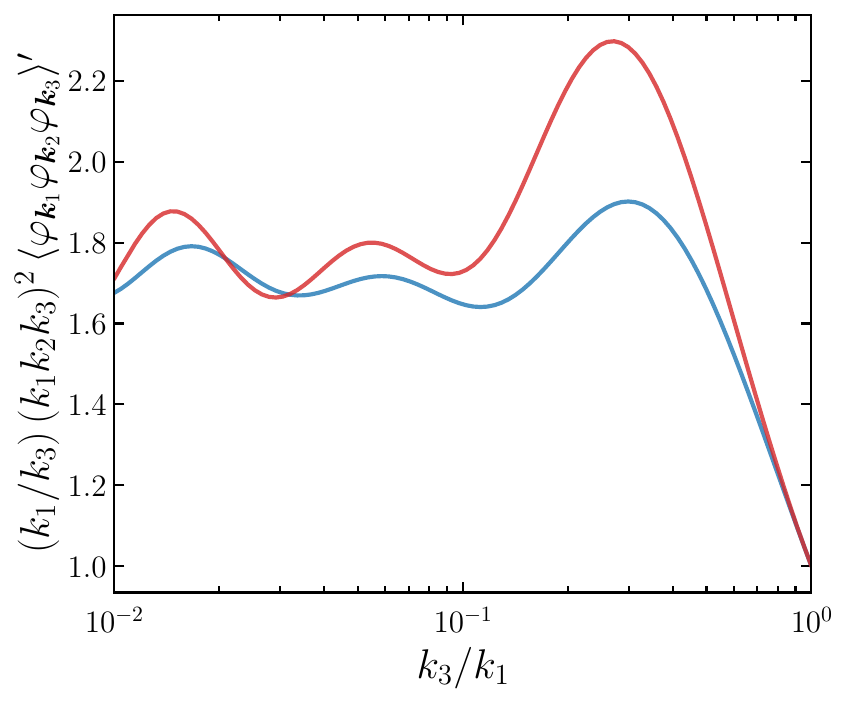}}
  \vspace*{0.2cm}
  \caption{Rescaled three-point correlator $(k_1/k_3)\times(k_1 k_2 k_3)^2\braket{\varphi_{\bm{k}_1} \varphi_{\bm{k}_2} \varphi_{\bm{k}_3}}'$ for the interaction $(\partial_i \varphi)^2\Psi$ (\textit{left}) and $\dot{\varphi}\Psi^2$ (\textit{right}) in the isosceles-triangle configuration $k_1=k_2$, for $c_s=0.1$ at weak mixing (\textcolor{pyblue}{blue}) and at strong mixing (\textcolor{pyred}{red}). For illustration purposes, we have normalised the amplitudes of the signals to be same in the equilateral configuration. The numerical parameters are $\Delta_r = 10^{-3}, n_{\text{disc}} = 10^4$ and $\Delta N = 5.5$. [\href{https://github.com/deniswerth/CosmoFlow/blob/main/CosmoFlow/PhiPsi/KinematicDependence.ipynb}{\faGithub}]}
  \label{fig: soft limits}
\end{figure}

\paragraph{Soft limits.} Scanning soft limits of cosmological correlators is computationally very expensive, as a large hierarchy of modes requires to resolve more massless $e$-folds in the sub-horizon regime, see Sec.~\ref{subsec: running time}. Therefore, these calculations are long and need parallelisation or additional computational capabilities, such as the use of HPC clusters. Nevertheless, for mildly soft configurations, we show here that soft limits can be computed in a reasonable amount of time ($\sim20$min) on a laptop. In Fig.~\ref{fig: soft limits}, we plot the rescaled three-point correlator---or equivalently the dimensionless bispectrum shape---in the isosceles-triangle configuration for two different cubic interactions, in the weakly and strongly mixed regimes. The corresponding notebook is \href{https://github.com/deniswerth/CosmoFlow/blob/main/CosmoFlow/PhiPsi/KinematicDependence.ipynb}{\faGithub}. The critical property appearing in these plots is the presence of oscillations whose frequency is set by the (effective) mass of the heavy field. These characteristic patterns are known as cosmological collider signals, and enable to identify new particles during inflation, see e.g.~\cite{Chen:2009we, Chen:2009zp, Noumi:2012vr, Arkani-Hamed:2015bza, Werth:2023pfl, Pinol:2023oux}. Our code is precise enough to resolve these signals without difficulty, even for double-exchange diagrams or at strong mixing for which analytical methods fail. Moreover, it allows us to reconstruct the history of cosmological collider signals as function of time, which enables us to have direct access to particle production during inflation. Movies of these cosmological collider flows can be found \href{https://github.com/deniswerth/Cosmological-Collider-Flow}{here}. Yet, for those willing to study soft limits with \textsf{CosmoFlow}, a note of caution is in order. One must be careful when parallelising the scanning because configurations close to equilateral configurations run faster than soft ones, as less massless $e$-folds are needed. Indeed, for soft configurations, one must resolve all the dynamics between the time when the long mode crosses the horizon and the time at which the short mode crosses its horizon (and of course one should add to it initial massless $e$-folds $\Delta N$ for a correct initialisation of the flow equations). Therefore, points for the soft limit discretisation should be equally distributed among the different cores. 

\vskip 4pt
It is interesting to note that soft limits usually allow for simplification when it comes to deriving analytical results, e.g.~by the use of operator product expansions and/or factorisation of nested time integrals~\cite{Arkani-Hamed:2015bza, Lee:2016vti, Pinol:2021aun, Tong:2021wai, Qin:2022lva, Qin:2023bjk}. However, these methods do not capture the entirety of the signal. For example, they only capture the so-called \textit{non-analytic} part of the signal, associated with particle production. Conversely, the cosmological flow approach allows for complete and exact results, and can be used to complement and/or guide analytical studies. In the end, predicting the theory, scale, or shape dependence of cosmological correlators poses an intricate challenge. However, through these examples, we have illustrated that \textsf{CosmoFlow} provides the means for achieving accurate predictions.

\subsection{Massive Fields in de Sitter and Triple-$H$ Integrals}

For generic theories in de Sitter, when computing cosmological correlators, although conceptually simple, the bulk time integrals typically cannot be evaluated analytically. In fact, cosmological correlators are complicated enough that, even at tree-level, their full analytical expressions usually contain a plethora of special and hypergeometric functions, some of which have not been yet studied. In this section, we illustrate how \textsf{CosmoFlow} enables one to probe some properties of these functions in physical kinematic regimes. 

\vskip 4pt
Let us consider the three-point function in momentum space of scalar operators in de Sitter with scaling dimension $\Delta$, at equal time $\tau_\star$. The dimension of the operator is related to the (bare) mass $m$ of the corresponding bulk scalar field by
\begin{equation}
    \Delta = \frac{3}{2} + i\mu\,, \quad \mu \equiv \sqrt{\left(\frac{m}{H}\right)^2 - \frac{9}{4}}\,.
\end{equation}
Formally, the solution has the following integral representation of three Hankel functions
\begin{equation}
    \begin{aligned}
    \braket{\O_{\bm{k}_1} \O_{\bm{k}_2} \O_{\bm{k}_3}}' &=  2\text{Re}\left(\mathcal{C}_\mu(\tau_\star, \{k_i\})\,\int_{-\infty}^{\tau_\star} \d \tau \tau^{1/2} H_{i\mu}^{(1)}(-k_1 \tau) H_{i\mu}^{(1)}(-k_2 \tau) H_{i\mu}^{(1)}(-k_3 \tau)\right) \\
    &\propto k_3^{3\Delta - 6}\, \left(\frac{k_1}{k_3}\right)^{2i\mu} \, \left(\frac{k_2}{k_3}\right)^{2i\mu} \, F_4\left(\alpha_-, \alpha_+; \gamma_-, \gamma_+; \frac{k_1^2}{k_3^2}, \frac{k_2^2}{k_3^2}\right) \,,
    \end{aligned}
\end{equation}
with $\alpha_\pm \equiv \tfrac{1}{2}(\tfrac{3}{2} \pm i\mu)$, $\gamma_\pm \equiv 1 \pm i\mu$, $\mathcal{C}_\mu(\tau_\star, \{k_i\}) \equiv \left(\tfrac{\pi}{4}\right)^3 e^{-3\pi\mu} (-\tau_\star)^{9/2} \,\Pi_{i=1}^3\, H_{-i\mu}^{(2)}(-k_i \tau_\star)$, and where $F_4$ is the Appell generalised hypergeometric function of two variables defined by the following double series
\begin{equation}
    F_4(\alpha_-, \alpha_+; \gamma_-, \gamma_+; x, y) = \sum_{n, m=0}^\infty \frac{(\alpha_-)_{n+m}\, (\alpha_+)_{n+m}}{(\gamma_-)_n\, (\gamma_+)_m\, n! m!}\, x^n y^m\,,
\end{equation}
for $\sqrt{|x|} + \sqrt{|y|} < 1$ and where $(\alpha)_n$ is the Pochhammer symbol. This function is also sometimes called Kampé de Fériet and some aspects have been studied in~\cite{Appell1926, Bateman1953}. Such integrals are usually referred to as triple-$H$ integrals, and are common in the CFT literature, see e.g.~\cite{Bzowski:2013sza} for a detailed analysis of conformal symmetry in momentum space where such integrals appear. This expression can be found as the unique solution to a pair of differential equations~\cite{Exton_1995, Prudnikov1992} that results from conformal Ward identities~\cite{Maldacena:2011nz, Creminelli:2012ed, Bzowski:2013sza}. However, written this way, the closed-form expression is not very illuminating. Moreover, note that this series representation is not that useful because it converges when $k_1 + k_2 < k_3$, i.e.~for \textit{unphysical} kinematic configurations.\footnote{Note that the \textsf{AppellF4} function in Mathematica is not useful in this case, and Python does not contain a vectorised version of the Hankel function of imaginary order, which complicates the numerical evaluation of such integrals.}

\begin{figure}[h!]
  \centering
  \subfloat{\includegraphics[width=0.45\textwidth]{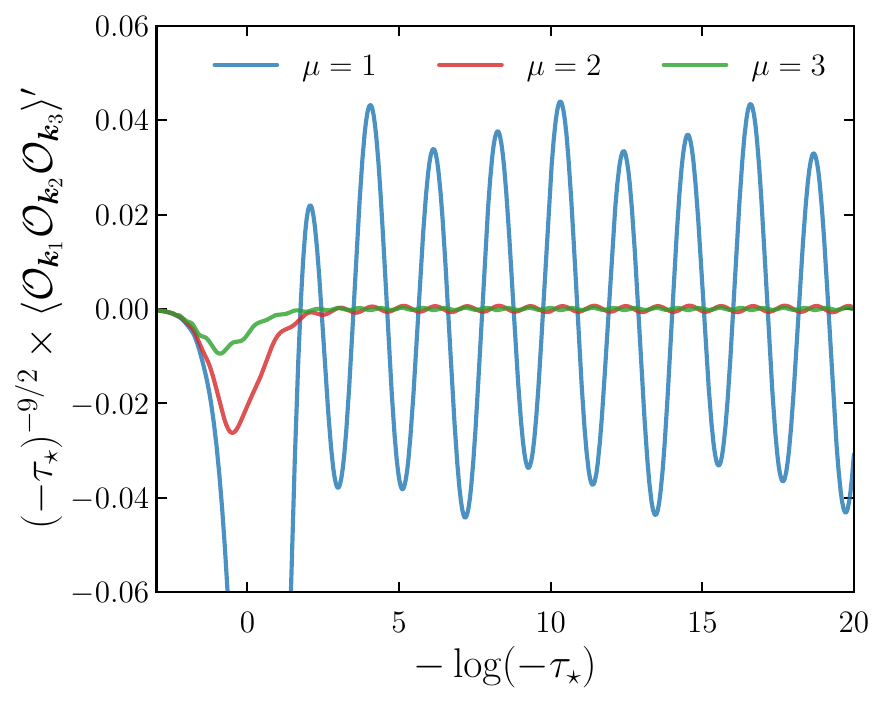}}
  \hspace*{0.2cm}
  \subfloat{\includegraphics[width=0.45\textwidth]{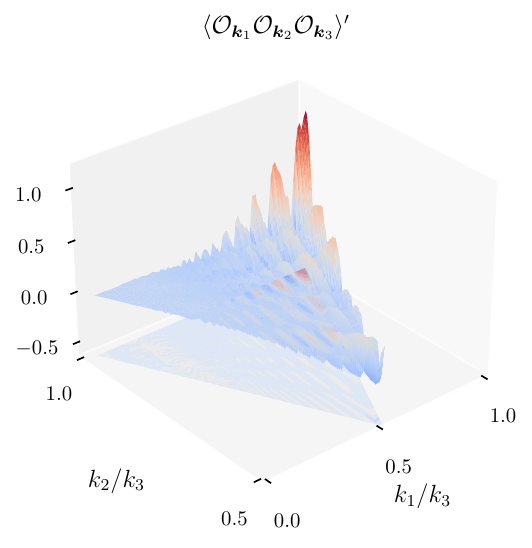}}
  \vspace*{0.2cm}
  \caption{\textit{Left}: Cosmological flow of the rescaled three-point function of $\O$ in the equilateral configuration $k_1=k_2=k_3=1$, varying the scaling dimension $\Delta = \tfrac{3}{2} + i\mu$. \textit{Right}: Rescaled three-point function $(-\tau_\star)^{9/2}(k_1 k_2 k_3)^2\,\braket{\O_{\bm{k}_1} \O_{\bm{k}_2} \O_{\bm{k}_3}}'$ in the entire physical kinematic space and normalised to unity in the equilateral configuration, as a function of the two dimensionless ratios $k_1/k_3$ and $k_2/k_3$, at the time $N_\star \equiv \log(-1/\tau_\star) = 5$ for $\mu = 2$. [\href{https://github.com/deniswerth/CosmoFlow/blob/main/CosmoFlow/TripleH/TripleH.ipynb}{\faGithub}]}
  \label{fig: Triple-H}
\end{figure}

\vskip 4pt
For the numerical study of this function, we use the fact that it is a solution of the three-point function of a scalar operator with dimension $\Delta$ and with a cubic self-interaction $\O^3$. The relevant tensors appearing in~(\ref{eq: Hamiltonian}) are $\Delta_{\O\O} = 1, M_{\O\O} = -(k^2/a^2 + m^2)$ and $A_{\O\O\O} = -2$. In Fig.~\ref{fig: Triple-H}, we both show the time evolution of the three-point function and the kinematic dependence in the entire physical space. The correlator has been rescaled because its late-time behaviour scales as $(-\tau_\star)^{-9/2}$. We clearly observe the
presence of several harmonics whose amplitudes are comparable, leading to a notable oscillatory
feature. These oscillations result from the interference between quasi-normal modes of the bulk scalar field mode functions. Indeed, in the late-time limit, it is instructive to expand the massive field mode function $f_k(\tau)$ in powers of $\tau^\Delta$ that correspond to quasi-normal modes in the static patch of de Sitter spacetime~\cite{Du:2004jt, Jafferis:2013qia}
\begin{equation}
    f_k(\tau) = \frac{H}{\sqrt{k^3}}\,\sum_{n=0}^{\infty}\sum_{\pm} c_n^\pm(\mu) \left(\frac{-k\tau}{2}\right)^{\frac{3}{2}+2n\pm i\mu}\,, \quad c_n^{\pm}(\mu) \equiv \pm \frac{(-1)^n}{n!}\frac{\sqrt{2\pi}}{\Gamma(n+1\pm i\mu)}\frac{e^{\pm\pi\mu/2}}{\sinh\pi\mu}\,.
\end{equation}
The quasi-normal modes correspond to $e^{-\omega_n^\pm t}$ with frequencies $\omega_n^\pm/H = \frac{3}{2}+2n\pm i\mu$, and $t$ being the cosmic time. The factor $3/2$ in the real part of $\omega_n^\pm$ reflects the dilution of the mode due to the expansion of the universe. The imaginary part of $\omega_n^\pm$ represents the positive and negative frequency oscillation of massive states. For a light field, the frequency $\omega_n^\pm$ is real, and the positive frequency is the dominant contribution. A notable feature of the amplitude of the quasi-normal modes is the factor $e^{\pm\pi\mu/2}$ which---when raised to the fourth power---corresponds to a Boltzmann suppression associated with the de Sitter temperature $T_{\text{dS}}=H/2\pi$.\footnote{The actual thermal factor for the two-point correlator is $e^{-\pi\mu}$ which corresponds to the transition amplitude for creating a pair of massive particles. One should take the square of it to get the probability of this event.} The three-point function is then a superposition of two modes: (i) one at a frequency $\omega/H = 3\mu$ coming from the superposition of three positive frequency quasi-normal modes, and (ii) one at frequency $\omega/H = \mu$ coming from the superposition of two positive and one
negative frequency quasi-normal modes. It is interesting to see that the late-time behaviour of the three-point function does not contain any local terms---i.e.~analytic in the momenta---as it oscillates around the origin. This is the result of precise cancellation between the leading quasi-normal modes. Using \textsf{CosmoFlow}, note that we cannot study analytic properties of special and hypergeometric functions, i.e.~analytically continue momenta to unphysical configurations.

\section{Performances}
\label{sec: Performances}

Lastly, in this section, we assess the performances of \textsf{CosmoFlow} and give new users some benchmarks for the expected integration time and accuracy. Throughout the tests we perform here, we set $g=1$ for the single-field theory~(\ref{eq: dphi3 action}), and $c_s=1, m/H = 2, \rho/H = 0.1$ and $\lambda_1 = 1$, i.e.~selecting the interaction $(\partial_i \varphi)^2\Psi$, for the two-field theory~(\ref{eq: phi-psi Lagrangian}).

\subsection{Running Time}
\label{subsec: running time}

We begin by estimating how the numerical runtime to integrate the flow equations scales with various parameters, e.g.~the number of massless $e$-folds in the sub-horizon regime, or the correlator kinematic configuration. All tests were performed on Macbook Pro with M1 CPU and running MacOS 11.6 Big Sur, and the computations can be found in the notebooks \href{https://github.com/deniswerth/CosmoFlow/blob/main/CosmoFlow/Massless_dphi3/Performances.ipynb}{\faGithub} and \href{https://github.com/deniswerth/CosmoFlow/blob/main/CosmoFlow/PhiPsi/Performances.ipynb}{\faGithub}, for the single-field theory~(\ref{eq: dphi3 action}) and the two-field theory~(\ref{eq: phi-psi Lagrangian}), respectively.

\begin{figure}[h!]
  \centering
  \includegraphics[width=0.6\textwidth]{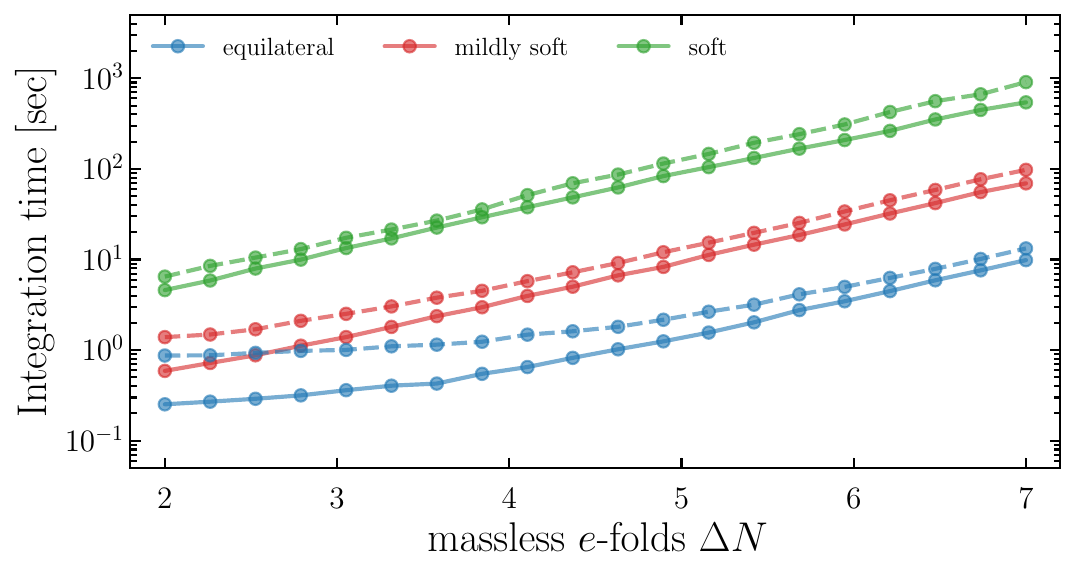}
  \vspace*{0.2cm}
  \caption{Integration time (in seconds) as a function of massless sub-horizon $e$-folds $\Delta N$ to initialise the flow equations for the three-point functions, in the equilateral configuration $k_1=k_2=k_3$ (\textcolor{pyblue}{blue}), mildly soft limit $k_3/k_1 = k_3/k_2 = 10^{-1}$ (\textcolor{pyred}{red}), and in the soft limit $k_3/k_1 = k_3/k_2 = 10^{-2}$ (\textcolor{pygreen}{green}), in the single-field theory~(\ref{eq: dphi3 action}) (solid line) and the two-field theory~(\ref{eq: phi-psi Lagrangian}) (dashed line), respectively. The benchmark parameters are $g=1$ for the single-field theory, and $c_s=1, m/H = 2, \rho/H = 0.1$ and $\lambda_1 = 1$ ($(\partial_i \varphi)^2\Psi$ interaction) for the two-field theory. Numerical parameters have been set to $\Delta_r = 10^{-3}$ and $n_{\text{disc}} = 10^4$. [\href{https://github.com/deniswerth/CosmoFlow/blob/main/CosmoFlow/PhiPsi/Performances.ipynb}{\faGithub}][\href{https://github.com/deniswerth/CosmoFlow/blob/main/CosmoFlow/Massless_dphi3/Performances.ipynb}{\faGithub}]}
  \label{fig: Performances massles efolds}
\end{figure}

\begin{figure}[h!]
  \centering
  \includegraphics[width=0.6\textwidth]{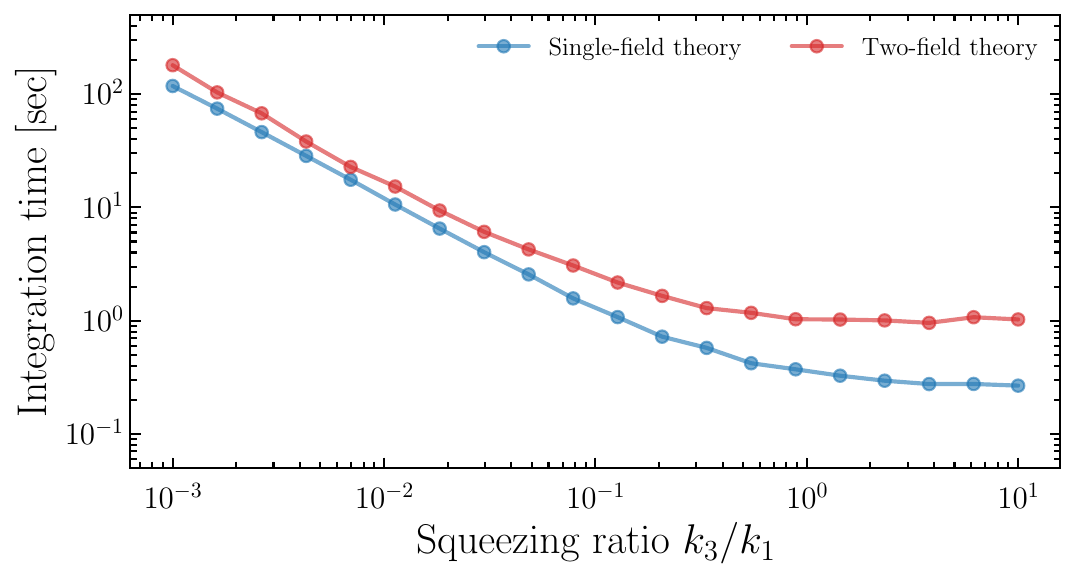}
  \vspace*{0.2cm}
  \caption{Integration time (in seconds) as a function of the dimensionless squeezing ratio $k_1/k_3$ for the three-point functions in the isosceles kinematic configuration $k_1=k_2$, in the single-field theory~(\ref{eq: dphi3 action}) (\textcolor{pyblue}{blue}) and the two-field theory~(\ref{eq: phi-psi Lagrangian}) (\textcolor{pyred}{red}), respectively. We have chosen $\Delta N = 3, \Delta_r = 10^{-3}$ and $n_{\text{disc}} = 10^4$, and $g=1$ for the single-field theory, and $c_s=1, m/H = 2, \rho/H = 0.1$ and $\lambda_1 = 1$ ($(\partial_i \varphi)^2\Psi$ interaction) for the two-field theory. Note that in the soft limit, more massless $e$-folds are needed to obtain accurate results, resulting in exponentially increasing the integration time. [\href{https://github.com/deniswerth/CosmoFlow/blob/main/CosmoFlow/PhiPsi/Performances.ipynb}{\faGithub}][\href{https://github.com/deniswerth/CosmoFlow/blob/main/CosmoFlow/Massless_dphi3/Performances.ipynb}{\faGithub}]}
  \label{fig: Performances soft limit}
\end{figure}

\vskip 4pt
As we have seen previously, setting up initial conditions for the flow equations---either analytical using the massless approximation or using the introduced numerical $i\epsilon$ prescription---requires to start integrating well inside the horizon (typically of the smallest wavenumber) in order to properly initialise the correlators in the correct vacuum state. Naturally, the earlier we initialise the integration, the more accurate the result is, as the initial state can only be reached asymptotically. However, as we have seen in Sec.~\ref{subsec: Possible Issues}, accurately integrating massless $e$-folds requires a good precision to exactly cancel positive and negative frequency modes. Therefore, integrating the sub-horizon regime is computationally very expensive. Choosing the correct number of massless $e$-folds $\Delta N$ is a balance between precision and speed, as it is usually the case with numerical methods. In Fig.~\ref{fig: Performances massles efolds}, we show the integration time as a function of the number of massless $e$-folds in the sub-horizon regime $\Delta N$, in different kinematic configurations, and for different theories. One can clearly observe that the integration time scales exponentially with $\Delta N$. The reason is explained in the insert of Sec.~\ref{subsec: Possible Issues}. A numerical fit gives the following empirical formula
\begin{equation}
    \Delta t \approx \frac{e^{\Delta N}}{100} \left(\frac{k_{\text{L}}}{k_{\text{S}}}\right)^{-1} [\text{sec}]\,,
\end{equation}
where $\Delta t$ is the numerical integration time in seconds, $k_\text{L}$ (res.~$k_\text{S}$) is some typical long (resp.~short) wavenumber. The ratio $k_{\text{L}}/k_{\text{S}}$ indicates the kinematic softness of the correlator. Here, we see that the kinematic configuration of the correlator, especially its soft limit, also plays a significant role in the integration time. Explicitly, we define the number of massless $e$-folds $\Delta N$ with respect to the longest scale, i.e.~the mode which crosses its horizon first. After fixing $k_3 = 1$ to be the longest mode (we choose this scale so that its horizon is located at $N=0$, therefore arbitrarily defining $a \equiv e^N$), the soft limit is reached by increasing the two remaining modes $k_3 \ll k_1\sim k_2$ that exit their horizon later. It means that for soft enough configurations, there is a long phase between horizon crossing of $k_3$ and that of $k_1$ (or $k_2$) in which the short modes behave as if they were massless, i.e.~in their vacuum state. This phase is exponentially expensive in terms of integration time, as is shown in Fig.~\ref{fig: Performances soft limit} where we depict the integration time as a function of the scaling ratio $k_3/k_1$ both for single- and two-field theories. Decreasing $k_3/k_1$ is the same as increasing $\Delta N$. Probing soft limits is therefore exponentially expensive. It is interesting to notice that there is no drastic change in the integration time as we increase the number of fields, at least from one to two. Indeed, these results show that the complexity comes from the sub-horizon regime and is almost not sensitive to the number of flow equations to integrate. We will see in the next section that a typical choice for $\Delta N$ lies in between $4$ and $5$ $e$-folds, depending on the complexity and kinematic configuration of the correlator. Note also that the final integration time has been set to $N = 20$. The dynamics on super-horizon scales is drastically simpler than that in the sub-horizon regime, and is solved almost trivially with our code. As a matter of fact, increasing the number of super-horizon number of $e$-folds does not affect the integration time.

\vskip 4pt
Even though all the presented examples in this paper can be launched on usual laptops, some applications, e.g.~probing extremely soft configurations with $k_3/k_1\sim 10^{-4}$, require the use of parallelisation to run \textsf{CosmoFlow} on multiple cores and on machines that are more powerful than laptops. When using the code for these cases, we have conducted experiments on parallelisation with up to 16 cores, and our observations revealed that the overall execution time typically stabilises when increasing the core count. This phenomenon can be attributed to a decrease in memory processing speed, impacting computation and communication across the various cores. It happened that some calculations created excess memory issues, forcing the code to stop abruptly. We believe this can be attributed to insufficient RAM and an excessively low absolute tolerance. In this case, we recommend employing machines with higher RAM capacity and adjusting the solver's absolute tolerance (e.g., setting $\Delta_a = 10^{-20}$) to prevent memory saturation caused by overly precise stored information. We invite potential users to contact the developers for more information and tips.

\subsection{Convergence}

\begin{figure}[h!]
  \centering
  \includegraphics[width=0.6\textwidth]{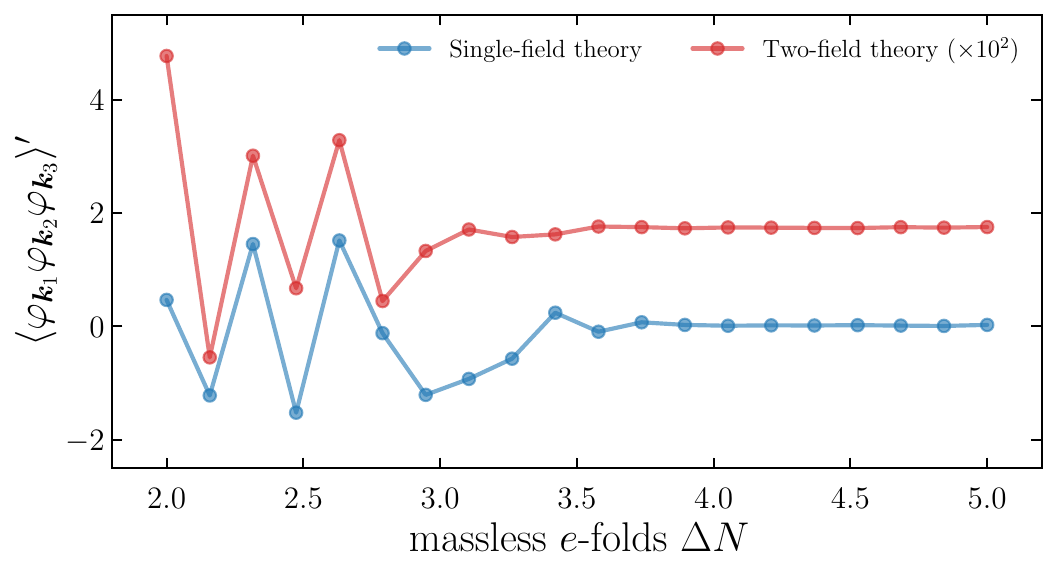}
  \vspace*{0.2cm}
  \caption{Three-point correlator $\braket{\varphi_{\bm{k}_1} \varphi_{\bm{k}_2} \varphi_{\bm{k}_3}}'$ as a function of massless sub-horizon $e$-folds $\Delta N$, in the equilateral configuration $k_1=k_2=k_3\equiv k$ for the single-field theory~(\ref{eq: dphi3 action}) (\textcolor{pyblue}{blue}) and the two-field theory~(\ref{eq: phi-psi Lagrangian}) (\textcolor{pyred}{red}), respectively. The scales have been chosen arbitrary by setting $k=1$. The benchmark parameters are $g=1$ for the single-field theory, and $c_s=1, m/H = 2, \rho/H = 0.1$ and $\lambda_1 = 1$ ($(\partial_i \varphi)^2\Psi$ interaction) for the two-field theory, with $\Delta_r = 10^{-4}$ and $n_{\text{disc}} = 10^4$. [\href{https://github.com/deniswerth/CosmoFlow/blob/main/CosmoFlow/PhiPsi/Performances.ipynb}{\faGithub}][\href{https://github.com/deniswerth/CosmoFlow/blob/main/CosmoFlow/Massless_dphi3/Performances.ipynb}{\faGithub}]}
  \label{fig: Convergence}
\end{figure}

In Fig.~\ref{fig: Convergence}, we show the final value of the three-point correlator $\braket{\varphi_{\bm{k}_1} \varphi_{\bm{k}_2} \varphi_{\bm{k}_3}}'$ in the equilateral configuration as a function of the number of massless $e$-folds in the sub-horizon regime, for the two theories we consider. We clearly see that the results start to converge around $\Delta N \approx 3.5$ and become stable. For the single-field theory for which
the exact analytical result is known, sub-5\% level of (relative) accuracy is
reached with fluctuations below 1\%. It is still interesting to notice that it seems that convergence is reached faster in the two-field theory than in the single-field theory, which is much simpler. Eventually, it is important to keep in mind that when increasing too much $\Delta N$, typically $\Delta N\geq 6$ in the equilateral configuration, the solution becomes unstable as more sub-horizon evolution requires more precision, see Sec.~\ref{subsec: Possible Issues}. In this case, one should also decrease the relative tolerance of the integrator $\Delta_r$.

\section{Conclusion}
\label{sec: Conclusion}

In this paper, we have introduced \textsf{CosmoFlow}, a numerical Python package for computing inflationary correlation functions that follows their cosmological flow. Given initial conditions in the asymptotic past and a fundamental theory for bulk degrees of freedom, correlators are uniquely determined by their time evolution. Our numerical implementation allows for a systematic and unified resolution of this flow, providing exact and fast results for correlators in regimes that were previously out of reach. We have deliberately chosen simpleness as our credo so that the code can be used, adapted and improved by a wide community. Our numerical implementation of the cosmological flow, the detailed user-guide, and the numerous tutorial notebooks to reproduce all the examples presented in this paper allow any new user to quickly become familiar with this approach and obtain immediate high-resolution results. The code is publicly available on \href{https://github.com/deniswerth/CosmoFlow}{GitHub}.

\vskip 4pt
Our work presents avenues for continued advancement, and we conclude by outlining key open directions, as well as possible extensions and improvements that remained to be explored. First, the current approach and its numerical implementation thus far only allow to systematically compute correlation functions of scalar degrees of freedom. However, for the sake of exhaustiveness, probing the physics during inflation also requires an in-depth understanding of the signatures of spinning particles. The cosmological flow formalism allows for a straight extension to include spinning fields. As for the numerical implementation, it requires keeping track of polarisation indices (which necessitates upgrading current objects to additional dimensions, some for field-space indices and others for polarisation states), properly encoding contractions of polarisation tensors with spatial momenta, and encompassing constraint equations within the flow equations. Second, the formalism also naturally allows one to compute higher-order correlators, starting from the four-point correlator. The relevant flow equations at tree-level have been derived, and it remains to be numerically implemented. However, solving higher-order flow equations would require more numerical power, simply because more equations need to be solved. Ideally, more sophisticated solvers need to be used (written in C++ with a Python wrapper to maintain simpleness), still avoiding hard-coding to make the code easy to modify or improve.

\vskip 4pt
The technical challenges encountered in investigating the earliest moments of our Universe has motivated us to develop the cosmological flow approach. In particle physics, where precision physics requires fast and accurate numerical codes to analyse and process a huge collection of data, the use of automated tools is a daily routine. However, the corresponding tools in cosmology are presently less advanced. Through the implementation of \textsf{CosmoFlow}, we have automated the process of deriving cosmological observables directly from fundamental theories. While this marks just the initial stride, it is evident that a substantial journey lies ahead, promising numerous new discoveries.

\paragraph{Acknowledgements.} We thank Diego Blas, David Mulryne, David Seery, Lukas Witkowski for helpful discussions. DW and SRP are supported by the European Research Council under the European Union’s Horizon 2020 research and innovation programme (grant agreement No 758792, Starting Grant project GEODESI). L.P. acknowledges funding support from the Initiative Physique des Infinis (IPI), a research training program of the Idex SUPER at Sorbonne Université. This article is distributed under the Creative Commons Attribution International Licence (\href{https://creativecommons.org/licenses/by/4.0/}{CC-BY 4.0}).

\newpage
\phantomsection
\addcontentsline{toc}{section}{References}
\small
\bibliographystyle{utphys}
\bibliography{references}

\end{document}